\documentclass{amsart}

\usepackage {epsfig}
\usepackage {graphicx}

 \newtheorem{thm}{Theorem}
 
 \newtheorem{lemma}[thm]{Lemma}
 
 \newtheorem{theorem}[thm]{Theorem}
 \theoremstyle{definition}
 
 \theoremstyle{remark}
 
 \newtheorem{question}[thm]{Question}
 \newtheorem{conjecture}[thm]{Conjecture}

 \newcommand{\Real}{\mathbb{R}}

 \newcommand{\Integer}{\mathbb{Z}}

\usepackage{graphicx}
\begin{document}

\title[Periodic waves in Theta neurons]
{Periodic travelling waves in the Theta model for synaptically
connected neurons}

\author{Guy Katriel}
\address{Einstein Institute of Mathematics, The Hebrew University of Jerusalem, Jerusalem, 91904, Israel}

\email{haggaik@wowmail.com}

\thanks{Partially supported by the Edmund Landau
Center for Research in Mathematical Analysis and Related Areas,
sponsored by the Minerva Foundation (Germany).}


\begin{abstract}
We study periodic travelling waves in the Theta model for a linear
continuum of synaptically-interacting neurons. We prove that when
the neurons are oscillatory, at least one periodic travelling of
every wave number always exists. In the case of excitable neurons,
we prove that no periodic travelling waves exist when the synaptic
coupling is weak, and at least two periodic travelling waves of
each wave-number, a `fast' one and a `slow' one, exist when the
synaptic coupling is sufficiently strong. We derive explicit upper
and lower bounds for the `critical' coupling strength as well as
for the wave velocities. We also study the limits of large
wave-number and of small wave-number, in which results which are
independent of the form of the synaptic-coupling kernel can be
obtained. Results of numerical computations of the periodic
travelling waves are also presented.

\end{abstract}

\maketitle

\section{introduction}
In this work we study periodic travelling waves in one-dimensional
continua of neurons described by the Theta model. The Theta model
\cite{ermentrout, ek, hibook, iz}, which is derived as a canonical
model for neurons near a `saddle-node on a limit cycle'
bifurcation, assumes the state of the neuron is given by an angle
$\theta$, with $\theta=(2l+1)\pi$, $l\in\Integer$ corresponding to
the `firing' state, and the dynamics described by
\begin{equation}\label{thn}
\frac{d\theta}{dt}=1-\cos(\theta)+(1+\cos(\theta))(\beta+I(t)),
\end{equation}
where $I(t)$ represents the inputs to the neuron. When $\beta<0$
this model describes an `excitable' neuron, which in the absence
of external input ($I\equiv 0$) approaches a rest state, while if
$\beta>0$ this represents an `oscillatory' neuron which performs
spontaneous oscillations in the absence of external input.

A model of synaptically connected neurons on a continuous spacial
domain $\Omega$ takes the form:
\begin{equation}\label{de}
\frac{\partial \theta (x,t)}{\partial t}=
1-\cos(\theta(x,t))+(1+\cos(\theta(x,t)))\Big[
\beta+g\int_{\Omega}{J(x-y)s(y,t)dy}\Big],
\end{equation}
\begin{equation}\label{syn}
\frac{\partial s (x,t)}{\partial t}+s(x,t)=P(\theta(x,t))(1-c s(x,t)),
\end{equation}
where $J$ - the synaptic-coupling kernel - is a positive function
and $P$ is defined by
\begin{equation}\label{defp}
P(\theta)=\sum_{l=-\infty}^{\infty}{\delta(\theta -(2l+1)\pi)}.
\end{equation}
Here $s(x,t)$ ($x\in\Omega, t\in\Real$) measures the synaptic
transmission from the neuron located at $x$, and according to
(\ref{syn}),(\ref{defp}) it decays exponentially, except when the
neuron fires ({\it{i.e.}} when $\theta(x,t)=(2l+1)\pi$,
$l\in\Integer$), when it experiences a jump. (\ref{de}) says that
the neurons are modelled as Theta-neurons, where the input
$I(x,t)$ to the neuron at $x$, as in (\ref{thn}), is given by
$$I(x,t)=g\int_{\Omega}{J(x-y)s(y,t)dy}.$$
$J(x-y)$ (here assumed to be positive) describes the relative
strength of the synaptic coupling from the neuron at $y$ to the
neuron at $x$, while $g$ is a parameter measuring the overall
coupling strength - we note that $g>0$ means that the connections
among neurons are excitatory while $g<0$ means that the
connections are inhibitory. It is naturally assumed that $J$
decays at infinity (precise assumptions will be presented later).

The above model, in the case $c>0$, is the one presented in
\cite{ermentrout, oe}. In the case $c=0$ this model is the one
presented in \cite{iz} (Remark 2) and \cite{rubin}. We always
assume $c\geq 0$.

When the geometry is linear, $\Omega=\Real$, it is natural to seek
travelling waves of activity along the line in which each neuron
makes one or more oscillations and then approaches rest, or even
where each neuron oscillates infinitely many times. In \cite{osan}
it was proven that for sufficiently strong synaptic coupling $g$,
at least two such waves, a slow and a fast one, exist, and also
that they always involve each neuron firing more than one time
before it approaches rest, while for sufficiently small $g$ such
waves do not exist. It was not determined how many times each
neuron fires before coming to rest, and it may even be that each
neuron fires infinitely many times. Some numerical results in the
case of a one and a two-dimensional geometry were obtained in
\cite{oe}. At least as far as can be gleaned from these
simulations, it seems likely that, starting with initial
conditions in which a small patch of tissue is depolarized, more
and more of the neurons start oscillating, so it may be that the
large-time behavior is described by a {\bf{periodic travelling
wave}}. This motivates the study of periodic travelling waves,
with infinitely many spikes. These are solutions of
(\ref{de}),(\ref{syn}) which take the form
\begin{equation}\label{tform}
\theta(x,t)=\phi(kx+\omega t),
\end{equation}
\begin{equation}\label{sform}
s(x,t)=r(kx+\omega t),
\end{equation}
where the functions $\phi$ and $r$ satisfy
\begin{equation}\label{bnp0}
\phi(z+2\pi)=\phi(z)+2\pi m\;\;\;\forall z\in\Real
\end{equation}
for some integer $m$, called the winding number, and
\begin{equation}\label{bnr0}
r(z+2\pi)=r(z)\;\;\;\forall z\in\Real.
\end{equation}
$k$ is the wave-number and $\omega$ is the frequency. The
wavelength of the travelling wave is given by
$$L=\frac{2\pi}{k},$$
the time-period is given by
$$T=\frac{2\pi}{\omega},$$
and the wave-velocity is given by
$$v=\frac{\omega}{k}.$$
In this formulation the spike-times of the neuron located at $x$
are the values $t_n$ for which $\phi(kx+\omega t_n)=(2l+1)\pi$,
where $l$ is an integer. It is thus not assumed that the
spike-times are equally spaced, but as we shall prove below (lemma
\ref{nspikes}) they form a periodic sequence in the sense that
$t_{n+m}=t_n+T$, so that the winding number $m$ is the number of
spike events for each neuron in the duration of a period $T$.

We note here that periodic travelling waves have already been
studied in the context of one-dimensional integrate-and-fire
neural networks \cite{bres,osan1}, in which case it was possible
to obtain quite explicit analytical results. The study of periodic
travelling waves in the Theta model presents further analytical
difficulties.

Substituting (\ref{tform}),(\ref{sform}) into
(\ref{de}),(\ref{syn}), and setting $z=kx+\omega t$, we obtain
\begin{equation}\label{phieq}
\omega
\phi'(z)=h(\phi(z))+gw(\phi(z))\frac{1}{k}\int_{-\infty}^{\infty}
{J\Big(\frac{1}{k}(z-u)\Big)r(u)du},
\end{equation}
where
\begin{equation}\label{spec}
h(\theta)=1-\cos(\theta)+\beta(1+\cos(\theta)),\;\;\;w(\theta)=1+\cos(\theta),
\end{equation}
and
\begin{equation}\label{requ}
\omega r'(z)+r(z)=P(\phi(z))(1-cr(z)).
\end{equation}
Periodic travelling waves thus correspond to solutions of
(\ref{phieq}),(\ref{requ}) satisfying (\ref{bnp0}),(\ref{bnr0}).
Let us note that, assuming that $r$ satisfies (\ref{bnr0}), the
integral appearing in (\ref{phieq}) is a $2\pi$-periodic function
of $z$, and we can rewrite this integral as
$$\frac{1}{k}\int_{-\infty}^{\infty}
{J\Big(\frac{1}{k}(z-u)\Big)r(u)du}=
\int_{0}^{2\pi}{J_k(z-u)r(u)du},$$ where $J_k$ is the periodic
function
\begin{equation}\label{defjk}
J_k(z)=\frac{1}{k}\sum_{l=-\infty}^{\infty}{J\Big(\frac{1}{k}(z-2\pi
l)\Big)},
\end{equation}
so that we can rewrite (\ref{phieq}) as
\begin{equation}\label{phieq1}
\omega \phi'(z)=h(\phi(z))+gw(\phi(z))\int_{0}^{2\pi}
{J_k(z-u)r(u)du}.
\end{equation}

Let us now note the important fact that the problem of solving
(\ref{phieq1}),(\ref{requ}),(\ref{bnp0}),(\ref{bnr0}) can also be
interpreted as that of searching for {\it{rotating waves}} for
(\ref{de}),(\ref{syn}) in the case that the spacial geometry is
given by $\Omega=S^1$, so the neurons are placed on a ring,
parametrized by $x\in \Real/2\pi\Integer$ and the equations are
(\ref{syn}) and
\begin{equation}\label{ge}
\frac{\partial \theta (x,t)}{\partial t}= h(\theta(x,t))+g
w(\theta(x,t))\int_{0}^{2\pi}{K(x-y)s(y,t)dy},
\end{equation}
where $K\in L^{\infty}(\Real)$ satisfies:
\begin{equation}\label{jpos}\inf_{x\in\Real}{K(x)}> 0,
\end{equation}
\begin{equation}\label{jper}
K(x+2\pi)=K(x)\;\;\;\forall x\in\Real,
\end{equation}
and the solutions satisfy the periodicity conditions
\begin{equation}\label{bnt}
\theta(x+2\pi,t)=\theta(x,t)+2\pi m\;\;\;\forall x,t\in\Real
\end{equation}
\begin{equation}\label{bns}
s(x+2\pi,t)=s(x,t)\;\;\;\forall x,t\in\Real.
\end{equation}
Rotating waves of (\ref{syn}) and (\ref{ge}) are solutions of the
form:
\begin{equation}\label{tw1}
\theta(x,t)=\phi(x+\omega t)
\end{equation}
\begin{equation}\label{tw2}
s(x,t)=r(x+ \omega t).
\end{equation}
with $\phi$,$r$ satisfying (\ref{bnp0}),(\ref{bnr0}). Substituting
(\ref{tw1}),(\ref{tw2}) into (\ref{syn}),(\ref{ge}), and setting
$z=x+\omega t$ we obtain
\begin{equation}\label{teq0}
\omega \phi'(z)= h(\phi(z))+ g
w(\phi(z))\int_{0}^{2\pi}{K(z-y)r(y)dy},
\end{equation}
and (\ref{requ}). (\ref{teq0}) is the same as (\ref{phieq1}), with
$K=J_k$. Thus studying the periodic travelling waves with wave
number $k$ in the case $\Omega=\Real$ is equivalent to studying
the rotating waves in the case $\Omega=S^1$, when we take $K=J_k$.
We shall obtain general results about rotating waves in the case
$\Omega=S^1$, {\it{i.e.}} about the problem
(\ref{teq0}),(\ref{requ}),(\ref{bnp0}),(\ref{bnr0}), to be solved
for $\omega,\phi(z),r(z)$. We note that in the ring
interpretation, $\omega$ is the {\it{velocity}} of the rotating
wave. From these general results about rotating waves on rings,
specialized to $K=J_k$, we will derive results about periodic
travelling waves of (\ref{de}),(\ref{syn}) in the case
$\Omega=\Real$.

In section \ref{pre} we show that in the case that the winding
number $m=0$, there can exist only trivial rotating waves. Thus
the interesting cases are when $m>0$. Here we study the case
$m=1$, the case $m>1$ being beyond our reach. Thus, this work
concentrates on the first non-trivial case, and we note also that
this is a case in which the spike-events are regularly-spaced.

Our central results about existence, nonexistence and multiplicity
of rotating waves on $\Omega=S^1$ can be summarized as follows
(see figures \ref{syn1},\ref{syn2} for the {\it{simplest}}
diagrams consistent with these results):

\begin{theorem}
\label{sum} Assume $K\in L^{\infty}(\Real)$ satisfies (\ref{jpos})
and (\ref{jper}).

\noindent (I) In the oscillatory case $\beta>0$: for all $g>0$
there exists a rotating wave, with velocity going to $+\infty$ as
$g\rightarrow+\infty$ and to $0$ as $g\rightarrow -\infty$.

\noindent (II) In the excitable case $\beta<0$: there exist
$g_{crit}>0$ such that

\noindent (i) For $g<g_{crit}$ there exist no rotating waves.

\noindent (ii)  for $g>g_{crit}$ there exist at least two rotating
waves, a `fast' and a `slow' one, in the sense that their
velocities approach $+\infty$ and $0$, respectively, as
$g\rightarrow +\infty$.

\noindent (III) In the boundary-case $\beta=0$ there exists a
rotating wave for any $g>0$, and no rotating wave when $g\leq 0$.
\end{theorem}

We thus see that while in the oscillatory case $\beta>0$ there
exists a rotating wave for any value of $g$ - even if $g<0$ so
that the coupling is inhibitory, in the excitable case existence
of rotating waves requires the coupling to be excitatory and of
sufficiently large magnitude, and in this case we have {\it{two}}
rotating waves.

Returning to the case of periodic travelling wave in the case
$\Omega=\Real$, we will see that the above theorem implies the
following theorem.

\begin{theorem}\label{pertra} Assume that $J$ satisfies
\begin{equation}\label{jdecay}
J,J'\in L^{1}(\Real)
\end{equation}
\begin{equation}\label{jpo}
J(z)>0 \;\;\;\forall z\in\Real
\end{equation}

\noindent Then:

\noindent (I) In the oscillatory case $\beta>0$: for any
wave-number $k>0$ there exists a periodic travelling wave with
wave number $k$ for any $g\in \Real$, with frequency going to
$+\infty$ as $g\rightarrow +\infty$ and to $0$ when $g\rightarrow
-\infty$.

\noindent (II) In the excitable case $\beta<0$: for any $k>0$
there exist $g_{crit}(k)>0$ such that

\noindent (A) (i) For $g<g_{crit}(k)$ there are no periodic
travelling waves with wave-number $k$.

\noindent (ii) For $g>g_{crit}(k)$ there are at least two periodic
travelling waves with wave-number $k$, with frequencies going to
$+\infty$ and to $0$ as $g\rightarrow +\infty$.

\noindent (B) There exist positive constants
$\underline{g}_{crit}$, $\overline{g}_{crit}$ such that
$$\underline{g}_{crit}\leq g_{crit}(k)\leq
\overline{g}_{crit}\;\;\;\forall k>0.$$

\noindent (III) In the boundary case $\beta=0$, for any $k>0$:

\noindent (i) For any $g>0$ there exists a periodic travelling
wave with wave number $k$ with frequency going to $+\infty$ as
$g\rightarrow \infty$.

\noindent (ii) For any $g\leq 0$ there are no periodic travelling
wave with wave-number $k$.
\end{theorem}

In particular part (II)(B) of the above theorem implies that when
$\beta<0$, for $g<\underline{g}_{crit}$ there exist no periodic
travelling waves of {\it{any}} wave-number, while for
$g>\overline{g}_{crit}$ there exist at least two periodic
travelling waves of {\it{every}} wave-number.

The above theorems will follow from more precise results which
will be proven in sections \ref{reduction}-\ref{genper}. In fact,
for given wave-number $k>0$, we will describe the curve $\Sigma_k$
in the $(\omega,g)$-plane, consisting of all pairs $(\omega,g)$
such that $\omega$ is the frequency of a travelling wave of
wave-number $k$ for coupling-strength $g$. We shall prove that
$\Sigma_k$ can be represented as the graph of a function
$g=g_k(\omega)$, and derive properties of the function $g_k$ which
will imply theorem \ref{pertra}, and more, including bounds for
the frequencies of the periodic travelling waves and for the
critical value $g_{crit}(k)$ in the case $\beta<0$.

We shall also obtain some more precise results in the limiting
cases $k\rightarrow \infty$ and $k\rightarrow 0$. We shall show
that for $k$ large, $J_k$ is nearly constant. It is thus pertinent
to study the problem
(\ref{teq0}),(\ref{requ}),(\ref{bnp0}),(\ref{bnr0}) in the case
that $K$ is constant, or in other words the question of rotating
waves on $\Omega=S^1$, when the coupling is uniform. Fortunately,
in this case, as we shall see in section \ref{constant}, the
problem simplifies considerably, and we can obtain more precise
results than those given in theorem \ref{sum} for the case of
general $K$, such as precise multiplicity results and closed
analytic expressions for the coupling-strength vs. wave-velocity
curves $\Sigma$, in an elementary fashion. In section \ref{large}
we shall see that these results imply results on periodic waves
with large wave number. The limit $k\rightarrow 0$ is studied in
section \ref{smallwn}.

The whole issue of stability, which we discuss briefly in section
\ref{stab}, remains quite open and awaits future investigation.

Although a central motivation for our study of rotating waves on
rings is obtaining results about periodic waves on a line, the
study of rotating waves on rings also has independent interest.
The model considered here, in the case $\beta<0$, describes waves
in an excitable medium, about which an extensive literature exists
(see \cite{winfree} and references therein). However, most models
consider diffusive rather than synaptic coupling. In the case of
the Theta model on a ring, with {\it{diffusive}} coupling, and
$m=1$, it is proven in \cite{er} that a rotating wave exists
regardless of the strength of coupling ({\it{i.e.}} the diffusion
coefficient), so that our results highlight the difference between
diffusive and synaptic coupling.

\section{preliminaries}
\label{pre}

We begin with an elementary calculus lemma which is useful in
several of our arguments below.
\begin{lemma}
\label{cut} Let $f:\Real\rightarrow\Real$ be a differentiable
function, and let $b,c\in \Real$, $b\neq 0$, be constants such
that we have the following property:
\begin{equation}
\label{kp}
f(z)=c \;\;\Rightarrow f'(z)=b.
\end{equation}
Then the equation $f(z)=c$ has at most one solution.
\end{lemma}

\noindent {\sc{proof:}} Assume by way of contradiction that the
equation $f(z)=c$ has at least two solutions $z_0<z_1$. Define
$S\subset \Real$ by
$$S=\{ z>z_0 \;|\; f(z)=c \}.$$
$S$ is nonempty because $z_1\in S$. Let $\underline{z}=\inf S$. By
continuity of $f$ we have $f(\underline{z})=c$. We have either
$\underline{z}>z_0$ or $\underline{z}=z_0$, and we shall show that
both of these possibilities lead to contradictions. If
$\underline{z}>z_0$, then by (\ref{kp}) we have
$$f(z_0)=f(\underline{z})=c$$
$$sign(f'(z_0))=sign(f'(\underline{z}))=sign(b)$$
so we conclude that there exists $z_2\in (z_0,\underline{z})$ with
$f(z_2)=c$, contradicting the definition of $\underline{z}$. If
$\underline{z}=z_0$ then $z_0$ is a limit-point of $S$, which
implies that $f'(z_0)=0$, contradicting (\ref{kp}) and the
assumption $b\neq 0$. These contradictions conclude our proof.

\vspace{0.4cm}

Turning now to our investigation of rotating waves in the case
$\Omega=S^1$, where $K\in L^{\infty}$ satisfies
(\ref{jpos}),(\ref{jper}), {\it{i.e.}} of solutions of
$\omega,\phi(z),r(z)$ of (\ref{teq0}),(\ref{requ}),
(\ref{bnp0}),(\ref{bnr0}), we note a few properties of the
functions $h(\theta)$ and $w(\theta)$ defined by (\ref{spec})
which will be used often in our arguments:
\begin{equation}\label{hpi}
h((2l+1)\pi)=2\;\;\;\forall l\in\Integer,
\end{equation}
\begin{equation}\label{hz}
h(2l\pi)=2\beta,\;\;\;\forall l\in\Integer,
\end{equation}
\begin{equation}\label{wz}
w((2l+1)\pi)=0\;\;\;\forall l\in\Integer,
\end{equation}
\begin{equation}\label{wnn}
w(2l\pi)=2\;\;\;\forall l\in\Integer,
\end{equation}

Let us first dispose of the case of zero-velocity waves,
$\omega=0$. We get the equations
\begin{equation}\label{teq00}
h(\phi(z))+ g w(\phi(z))\int_{0}^{2\pi}{K(z-y)r(y)dy}=0,
\end{equation}
\begin{equation}\label{req00}
r(z)= P(\phi(z))(1-c r(z)).
\end{equation}
If there exists some $z_0\in\Real$ with $\phi(z_0)=(2l+1)\pi$,
$l\in\Integer$, then, substituting $z=z_0$ into (\ref{teq00}) and
using (\ref{hpi}),(\ref{wz}), we obtain $2=0$, a contradiction.
Hence we must have
\begin{equation}\label{npi}
\phi(z)\neq (2l+1)\pi \;\;\;\forall z\in\Real,\;l\in\Integer
\end{equation}
which implies that $P(\phi(z))\equiv 0$, so that (\ref{req00})
gives $r(z)\equiv 0$, and (\ref{teq00}) reduces to
$h(\phi(z))\equiv 0$, and thus $\phi(z)$ is a constant function,
the constant being a root of $h(\theta)$. This implies, first of
all, that the winding number $m$ is $0$, since a constant
$\phi(z)$ cannot satisfy (\ref{bnp0}) otherwise. In addition the
function $h(\theta)$ must vanish somewhere, which is equivalent to
the condition $\beta\leq 0$. We have thus proven
\begin{lemma}
\label{zvel} Zero-velocity waves exist if and only if $m=0$ and
$\beta\leq 0$, and in this case they are just the stationary
solutions
$$r(z)\equiv 0$$
$$\phi(z)\equiv \pm \cos^{-1}\Big(\frac{\beta+1}{\beta-1}\Big)+2\pi l,
\;\;l\in\Integer.$$
\end{lemma}

Having found all zero-velocity waves, we may now assume
$\omega\neq 0$, so that our equations (\ref{teq0}),(\ref{requ})
for the rotating waves can be rewritten
\begin{equation}\label{teq}
\phi'(z)=\frac{1}{\omega}\Big[h(\phi(z))+ g
w(\phi(z))\int_{0}^{2\pi}{K(z-y)r(y)dy}\Big],
\end{equation}
\begin{equation}\label{req}
r'(z)+\frac{1}{\omega} r(z)=\frac{1}{\omega} P(\phi(z))(1-c r(z)),
\end{equation}
and rotating waves with non-zero velocity correspond to solutions
$\omega,r(z),\phi(z)$ of
(\ref{teq}),(\ref{req}),(\ref{bnp0}),(\ref{bnr0}).

\begin{lemma}\label{uniques}
Assume that $m\geq 1$ and $\omega,\phi(z),r(z)$ solve
(\ref{teq}),(\ref{req}),(\ref{bnp0}),(\ref{bnr0}). Then for each
$l\in \Integer$ there exists a {\it{unique}} solution of the
equation
\begin{equation}
\label{bas} \phi(z)=(2l+1)\pi.
\end{equation}
\end{lemma}

\noindent {\sc{proof:}} Since by (\ref{bnp0}) and the assumption
that $m\geq 1$ we have $\phi(\Real)=\Real$, the existence of
solutions of (\ref{bas}) is obvious. We note now the key fact
that, by (\ref{teq}),(\ref{hpi}) and (\ref{wz}),
\begin{equation}
\label{kf} \phi(z)=(2l+1)\pi,\;l\in\Integer\;\; \Rightarrow
\;\;\phi'(z)=\frac{2}{\omega}.
\end{equation}
so that lemma \ref{cut} implies the uniqueness.

\begin{lemma}\label{nspikes}
Assume $\omega,\phi(z),r(z)$ solve
(\ref{teq}),(\ref{req}),(\ref{bnp0}),(\ref{bnr0}). Then for each
interval of the form $[a,a+2\pi)$ there are precisely $m$ values
of $z\in [a,a+2\pi)$ for which $\phi(z)=(2l+1)\pi$ for some $l\in
\Integer$.
\end{lemma}

\noindent {\sc{proof:}} In the case $m=0$: we claim (\ref{npi})
must hold, which implies the statement of the lemma. Assume by way
of contradiction that $\phi(z_0)=(2l+1)\pi$ for some integer $l$.
By (\ref{kf}) and lemma \ref{cut} the equation $\phi(z)=(2l+1)\pi$
has at most one solution, contradicting the fact that, by
(\ref{bnp00}), we have $\phi(z_0)=\phi(z_0+2\pi)=(2l+1)\pi$.

Now assume $m\geq 1$. We define
$$l_0=\min \{ l\in\Integer \;|\; (2l+1)\pi\geq \phi(a)\} ,$$
$$l_1=\max \{ l\in\Integer \;|\; (2l+1)\pi<\phi(a+2\pi)\},$$
By (\ref{bnp0}), we have $\phi(a+2\pi)-\phi(a)=2\pi m$, so that
\begin{equation}\label{star}
l_1-l_0=m-1.
\end{equation}
By the intermediate-value theorem there exist solutions $z\in
[a,a+2\pi)$ of the equation (\ref{bas}) for all integers $l_0\leq
l\leq l_1$. By lemma \ref{uniques} these solutions are unique, and
we denote them by $z_l$, $l_0\leq l\leq l_1$. To conclude the
proof we only need to show that if $l<l_0$ or $l>l_1$ there exists
no solution $z\in [a,a+2\pi)$ of (\ref{bas}). If we assume by way
of contradiction that there exists $l<l_0$ with such that
(\ref{bas}) has a solution in $[a,a+2\pi)$, then by continuity
there exists $\overline{z}\in [a,a+2\pi)$ with
$\phi(\overline{z})=(2(l_0-1)+1)\pi$. Then, using (\ref{star}), we
have
$$\phi(\overline{z}+2\pi)=\phi(\overline{z})+2\pi
m=(2(l_0-1+m)+1)\pi=(2l_1+1)\pi.$$
On the other hand we have
$z_{l_1}\in [a,a+2\pi)$ with $\phi(z_{l_1})=(2l_1+1)\pi$. But
$\overline{z}+2\pi\neq z_{l_1}$ since $\overline{z}+2\pi>a+2\pi$,
so we have a contradiction to the uniqueness given by lemma
\ref{uniques}. An analogous argument can be made in the case
$l>l_1$, completing our proof. \vspace{0.4cm}

Let us note that if we knew that for rotating waves the function
$\phi(z)$ must be monotone, then lemma \ref{nspikes} would follow
immediately from (\ref{bnp0}).

\begin{question}
Is it true in general that rotating wave solutions are monotone
when $m\geq 1$?
\end{question}
\vspace{0.4cm}

We will now show that the trivial ``waves" of lemma \ref{zvel} are
the only ones that occur for $m=0$.

\begin{lemma}
\label{mz}
Assume $m=0$.

\noindent (i) If $\beta>0$ there are no rotating waves.

\noindent (ii) If $\beta\leq 0$ the only rotating waves are those
 given by lemma \ref{zvel}.
\end{lemma}

\noindent {\sc{proof:}} Assume $\omega,\phi(z),r(z)$ is a solution
of (\ref{teq0}),(\ref{requ}) satisfying (\ref{bnp0}) with $m=0$,
{\it{i.e.}}
\begin{equation}\label{bnp00}
\phi(z+2\pi)=\phi(z) \;\;\;\forall z\in\Real,
\end{equation}
and (\ref{bnr0}). We also assume $\omega\neq 0$, otherwise we are
back to lemma \ref{zvel}. We shall prove below that $\phi(z)$ must
satisfy (\ref{npi}), and hence that $P(\phi(z))\equiv0$, so that
by (\ref{requ}),(\ref{bnr0}) we have $r(z)\equiv 0$, so that
(\ref{teq0}) reduces to
\begin{equation}
\label{red} \omega\phi'(z)=h(\phi(z)).
\end{equation}
Since $\omega\neq 0$, if $h(\theta)$ has no roots ($\beta>0$),
(\ref{red}) has no solutions satisfying (\ref{bnp00}). If
$h(\theta)$ does have roots ($\beta\leq 0$) then the only
solutions of (\ref{red}) satisfying (\ref{bnp00}) are constant
functions, the constant being a root of $h(\theta)$, and we are
back to the same solutions given in lemma \ref{zvel}, which indeed
can be considered as rotating waves with arbitrary velocity.
\vspace{0.4cm}

Having found all possible rotating waves in the case $m=0$, we can
now turn to the case $m>0$. In fact, as was mentioned in the
introduction, we shall treat the case $m=1$, the cases $m>1$ being
harder. By lemma \ref{zvel} we know that there are no
zero-velocity waves, so we can assume $v\neq 0$ and define with
periodic conditions
\begin{equation}\label{bnp}
\phi(z+2\pi)=\phi(z)+2\pi \;\;\;\forall z\in\Real,
\end{equation}
\begin{equation}\label{bnr}
r(z+2\pi)=r(z)\;\;\;\forall z\in\Real.
\end{equation}

\section{Rotating waves on a ring: reduction to a one-dimensional equation}
\label{reduction}

Our object is to study the equations (\ref{teq}),(\ref{req}) for
$\omega,\phi(z),r(z)$ with periodic conditions
(\ref{bnp}),(\ref{bnr}). We will derive a scalar equation (see
(\ref{tt}) below) so that rotating waves are in one-to-one
correspondence with solutions of that equation.

We note first that, since by (\ref{bnp}) we have
$\phi(\Real)=\Real$, and since any rotating wave generates a
family of other rotating waves by translations, we may, without
loss of generality, fix
\begin{equation}\label{fix}
\phi(0)=\pi.
\end{equation}

\begin{lemma}
\label{one} Assume $\omega,\phi(z),r(z)$ satisfy
(\ref{teq}),(\ref{req}) with conditions
(\ref{bnp}),(\ref{bnr}),(\ref{fix}). Then
\begin{equation}
\label{one1}
z\in (0,2\pi) \Rightarrow \pi<\phi(z)<3\pi
\end{equation}
\end{lemma}

\noindent {\sc{proof:}} Lemma \ref{uniques} implies that the
equation $\phi(z)=(2l+1)\pi$ exactly one solution for each
$l\in\Integer$. In particular, since $\phi(0)=\pi$,
$\phi(2\pi)=3\pi$ we have $\phi(z)\neq\pi,3\pi$ for $z\in
(0,2\pi)$, and by continuity of $\phi(z)$ this implies
(\ref{one1}).

\begin{lemma}
\label{lampos} Assume $\omega,\phi(z),r(z)$ satisfy
(\ref{teq}),(\ref{req}) with conditions
(\ref{bnp}),(\ref{bnr}),(\ref{fix}). Then $\omega>0$, and
\begin{equation}\label{phit}
\phi'(0)=\frac{2}{\omega}.
\end{equation}
\end{lemma}

In other words the waves rotate clockwise. Of course in the
symmetric case $m=-1$ the waves will rotate counter-clockwise.

\vspace{0.4cm} \noindent {\sc{proof:}} By (\ref{fix}) and
(\ref{kf}) we have (\ref{phit}).  If $\omega$ were negative, then
$\phi$ would be decreasing near $z=0$, so for small $z>0$ we would
have $\phi(z)<\pi$, contradicting (\ref{one1}).

\vspace{0.4cm} Our next step is to solve (\ref{req}),(\ref{bnr})
for $r(z)$, in terms of $\phi(z)$. We will use the following
important consequence of lemma \ref{one}:

\begin{lemma}\label{ret}
$$P(\phi(z))|_{(-2\pi,2\pi)}=\frac{\omega}{2}\delta(z)$$
\end{lemma}

\noindent
{\sc{proof:}} By lemma \ref{one} we have
$$P(\phi(z))|_{(-2\pi,2\pi)}=\delta(\phi(z)-\pi)),$$
so we will show that
\begin{equation}
\label{jon} \delta(\phi(z)-\pi))=\frac{\omega}{2}\delta(z).
\end{equation}
Let $\chi\in C_0^{\infty}(\Real)$ be a test function. Using lemma
\ref{one} again we have
\begin{equation}
\label{arg3}
\int_{0}^{2\pi}{\chi(u)\delta(\phi(u)-\pi)du}=\int_{-\epsilon}^{\epsilon}{\chi(u)\delta(\phi(u)-\pi)du},
\end{equation}
where $\epsilon>0$ is arbitrary. In particular, since by lemma
\ref{lampos} $\phi'(0)=\frac{2}{\omega}>0$, we may choose
$\epsilon>0$ sufficiently small so that $\phi'(z)>0$ for
$z\in(-\epsilon,\epsilon)$, so that we can make a change of
variables $\varphi=\phi(u)$, obtaining, using (\ref{phit}),
\begin{eqnarray}\label{arg4}
\int_{-\epsilon}^{\epsilon}{\chi(u)
\delta(\phi(u)-\pi)du}&=&\int_{\phi(-\epsilon)}^{\phi(\epsilon)}
{\chi(\phi^{-1}(\varphi))\delta(\varphi-\pi)\frac{d\varphi}{\phi'(\phi^{-1}(\varphi))}}
\nonumber\\&=&\frac{\chi(0)}{\phi'(\phi^{-1}(\pi))}=
\frac{\chi(0)}{\phi'(0)}=\frac{\omega}{2}\chi(0).
\end{eqnarray}
This proves (\ref{jon}), completing the proof of the lemma.

\vspace{0.4cm}
By lemma \ref{ret} we can rewrite equation (\ref{req}) on the
interval $(-2\pi,2\pi)$ as
\begin{equation}\label{req1}
r'(z)+\Big(\frac{1}{\omega}+\frac{c}{2}\delta(z)\Big)
r(z)=\frac{1}{2} \delta(z),
\end{equation}
The solution of which is given by
\begin{equation}\label{gs}
r(z)=\Big(\frac{1}{2}H(z)+r(-\pi)e^{-\frac{\pi}{\omega}}\Big)e^{-(\frac{1}{\omega}
z +\frac{c}{2}H(z))},\;\;\;0<|z|<2\pi
\end{equation}
where $H$ is the Heaviside function: $H(z)=0$ for $z<0$, $H(z)=1$
for $z>0$.
Substituting $z=\pi$ into (\ref{gs}) and using (\ref{bnr}), we
obtain an equation for $r(-\pi)$ whose solution is
$$r(-\pi)=\frac{1}{2}(e^{\frac{\pi}{\omega}+\frac{c}{2}}-e^{-\frac{\pi}{\omega}})^{-1},$$
and substituting this back into (\ref{gs}), we obtain that the
solution of (\ref{req}),(\ref{bnr}) which we denote by
$r_{\omega}(z)$ in order to emphasize the dependence on the
parameter $\omega$, is given on the interval $(0,2\pi)$ by
\begin{equation}\label{fr}
r_{\omega}(z)=\Upsilon(\omega)e^{-\frac{1}{\omega}z} \;\;\;\;
0<z<2\pi,
\end{equation}
where
\begin{equation}\label{defups}
\Upsilon(\omega)\equiv
\frac{1}{2}\frac{e^{\frac{2\pi}{\omega}}}{e^{\frac{2\pi}{\omega}+\frac{c}{2}}-1}.
\end{equation}
We note that, for general $z\in\Real$, $r_{\omega}(z)$ is given as
the $2\pi$-periodic extension of the function defined by
(\ref{fr}) from $[0,2\pi]$ to the whole real line.

The following results, which can be computed from (\ref{fr}), will
be needed later
\begin{lemma}\label{propla}
We have for all $\omega>0$,
\begin{equation}
\label{mv}
\int_{0}^{2\pi}{r_{\omega}(u)du}=\frac{\omega}{2}\rho_c(\omega),
\end{equation}
where
$$\rho_c(\omega)=\frac{e^{\frac{2\pi}{\omega}}-1}{e^{\frac{2\pi}{\omega}+\frac{c}{2}}-1}.$$
\begin{equation}\label{minr}
\inf_{x\in\Real}{r_{\omega}(x)}=\Upsilon(\omega)e^{-\frac{2\pi}{\omega}},
\end{equation}
\begin{equation}\label{maxr}
\sup_{x\in\Real}{r_{\omega}(x)}=\Upsilon(\omega).
\end{equation}
\end{lemma}

We note that
\begin{equation}
\label{rho0} \rho_0(\omega)\equiv 1,
\end{equation}
a fact that considerably simplifies the formulas in the case
$c=0$. We note also that since
$$\lim_{\omega\rightarrow 0}{\rho_c(\omega)}=e^{-\frac{c}{2}},$$
and $\rho_c(\omega)$ is a monotone decreasing function when
$c>0$, we have
\begin{equation}\label{inrho}
0<\rho_c(\omega)< e^{-\frac{c}{2}}\;\;\;\forall \omega>0.
\end{equation}
\vspace{0.4cm}

The rotating waves thus correspond to solutions $\omega,\phi(z)$
of the equation
\begin{equation}\label{twe0}
\phi'(z)=\frac{1}{\omega}\Big[ h(\phi(z))+ g
w(\phi(z))\int_{0}^{2\pi}{K(z-y)r_{\omega}(y)dy}\Big],
\end{equation}
with $\phi(z)$ satisfying (\ref{fix}) and
\begin{equation}\label{bphi}
\phi(2\pi)=3\pi.
\end{equation}
To simplify notation, we define
\begin{equation}\label{defrl}
R_{\omega}(z)=\int_{0}^{2\pi}{K(z-y)r_{\omega}(y)dy},
\end{equation}
so that (\ref{twe0}) is rewritten as
\begin{equation}\label{twe}
\phi'(z)=\frac{1}{\omega}[h(\phi(z))+ g R_{\omega}(z) w(\phi(z))].
\end{equation}

We note that (\ref{twe}) is a nonautonomous differential equation
for $\phi(z)$, and since the nonlinearities are bounded and
Lipschitzian, the initial value problem (\ref{twe}),(\ref{fix})
has a unique solution, which we denote by $\phi_{\omega,g}$ (we
note that $\phi_{\omega,g}$ depends also on the parameters $\beta$
and $c$, but we shall suppress this dependence in our notation,
considering these parameters as fixed).

Rotating waves thus correspond to solutions $\omega>0$ of the
equation
\begin{equation}\label{ttt}
\phi_{\omega,g}(2\pi)=3\pi.
\end{equation}
Rewriting (\ref{twe}) and (\ref{fix}) we have
\begin{equation}\label{tww}
\phi_{\omega,g}'(z)=\frac{1}{\omega}[h(\phi_{\omega,g}(z))+g
R_{\omega}(z) w(\phi_{\omega,g}(z))],
\end{equation}
\begin{equation}\label{ip}
\phi_{\omega,g}(0)=\pi,
\end{equation}
and defining
\begin{equation}\label{aps}
\Psi(\omega,g)=\frac{1}{3\pi}\phi_{\omega,g}(2\pi),
\end{equation}
we obtain that rotating waves correspond to solutions $\omega>0$
of the equation
\begin{equation}\label{tt}
\Psi(\omega,g)=1
\end{equation}
where $g$ is considered a parameter in the equation (\ref{tt}). It
will be useful to define the synaptic-strength vs. velocity curve
\begin{equation}\label{defsig}
\Sigma=\{ (\omega,g)\;|\;\Psi(\omega,g)=1\},
\end{equation}
whose properties will be the central subject of study. Note that
the intersections of this curve with the line $g=g_0$ corresponds
to the rotating waves for synaptic strength $g_0$. We also note
that for each $(\omega,g)\in \Sigma$, there is a {\it{unique}}
rotating wave up to time-translations, by the uniqueness of the
solution of the initial-value problem (\ref{tww}),(\ref{ip}).

\vspace{0.4cm} We define the following quantities which will be
useful in the sequel:
\begin{equation}\label{defrlam}\overline{R}_{\omega}=\sup_{x\in\Real}{R_{\omega}(x)},\;\;\;
\underline{R}_{\omega}=\inf_{x\in\Real}{R_{\omega}(x)},\end{equation}
\begin{equation}\label{defouk}\overline{K}=\sup_{x\in\Real}{K(x)},\;\;\;
\underline{K}=\inf_{x\in\Real}{K(x)},\end{equation}
\begin{equation}\label{defk1}\|K\|_{L^1[0,2\pi]}=\int_0^{2\pi}{K(x)dx}.\end{equation}
We note that $\overline{K}<\infty$ since $K\in L^{\infty}(\Real)$.
By (\ref{jpos}), we always have $\underline{K}>0$.

The following bounds on
$\underline{R}_{\omega},\overline{R}_{\omega}$, will be useful.
The first follows immediately from (\ref{defrl}) and (\ref{mv}):
\begin{lemma}
\label{ulb}
\begin{equation}\label{bndrl}\frac{\omega}{2}\rho_c(\omega)\underline{K}\leq
\underline{R}_{\omega}\leq
\overline{R}_{\omega}\leq\frac{\omega}{2}\rho_c(\omega)\overline{K}.
\end{equation}
\end{lemma}

The next follows immediately from
(\ref{defrl}),(\ref{minr}),(\ref{maxr}):
\begin{lemma}\label{ulb1}
\begin{equation}\label{bndr2}\frac{1}{2}\frac{1}{e^{\frac{2\pi}{\omega}+\frac{c}{2}}-1}\|K\|_{L^1[0,2\pi]}\leq
\underline{R}_{\omega}\leq \overline{R}_{\omega}\leq
\frac{1}{2}\frac{e^{\frac{2\pi}{\omega}}}{e^{\frac{2\pi}{\omega}+\frac{c}{2}}-1}\|K\|_{L^1[0,2\pi]}
\end{equation}
\end{lemma}

For future use, we now restate the result of lemma \ref{one}:

\begin{lemma}
\label{b1} If $(\omega,g)\in\Sigma$ then
\begin{equation}
\label{one11} \pi<\phi_{\omega,g}(z)<3\pi \;\;\;\forall z\in
(0,2\pi).
\end{equation}
\end{lemma}

\section{Rotating waves on a ring: the case of uniform coupling}
\label{constant}

Assuming that the coupling in (\ref{teq}) is $K(x)\equiv K_0$ we
shall be able to solve for the rotating waves explicitly. In this
case we have, from (\ref{defrl}),(\ref{mv})
$$R_{\omega}(z)=K_0\int_{0}^{2\pi}{r_{\omega}(y)dy}=
\frac{K_0\omega}{2}\rho_c(\omega),$$ so that (\ref{tww}) reduces
to
\begin{equation}\label{tww0}
\phi_{\omega,g}'(z)=\frac{1}{\omega} h(\phi_{\omega,g}(z))+
\frac{K_0g}{2} \rho_c(\omega) w(\phi_{\omega,g}(z)) .
\end{equation}
The fact that (\ref{tww0}) is an autonomous equation is what makes
the treatment of the case where $K(x)$ is constant much simpler.
Indeed, assume that (\ref{tt}) holds, so that
\begin{equation}\label{sta}
\phi_{\omega,g}(2\pi)=3\pi.
\end{equation}
Then we have, using (\ref{tww0}), making a change of variables
$\varphi=\phi_{\omega,g}(z)$, and using (\ref{sta})
\begin{eqnarray}\label{lam}
2\pi&=&\int_{\pi}^{3\pi}{\frac{\phi_{\omega,g}'(z)
dz}{\frac{1}{\omega} h(\phi_{\omega,g}(z))+ \frac{K_0
g}{2}\rho_c(\omega)
w(\phi_{\omega,g}(z))}}\nonumber\\
&=&\int_{\phi_{\omega}(0)}^{\phi_{\omega}(2\pi)}{\frac{d\varphi}{\frac{1}{\omega}
h(\varphi)+ \frac{K_0 g}{2}\rho_c(\omega) w(\varphi)}}=
\int_{\pi}^{3\pi}{\frac{d\varphi}{\frac{1}{\omega} h(\varphi)+
\frac{K_0 g}{2}\rho_c(\omega) w(\varphi)}}.
\end{eqnarray}
Substituting the explicit expressions for $h$ and $w$ from (\ref{spec}),
and using the formula
\begin{equation}
\label{form}
\frac{1}{2\pi}\int_{\pi}^{3\pi}{\frac{d\phi}{A+Bcos(\phi)}}=\frac{1}{\sqrt{A^2-B^2}}
\;\;\;\;(|A|>|B|),
\end{equation}
(\ref{lam}) becomes
\begin{equation}\label{eqc}
1=\sqrt{4\Big(\frac{1}{\omega}\Big)^2\beta+2K_0g\frac{1}{\omega}\rho_c(\omega)},
\end{equation}
so that rotating waves correspond to solutions of (\ref{eqc}),
with their velocities given by $\omega$. We can rewrite
(\ref{eqc}) as
\begin{equation}\label{eqc1}
f_{c,\beta}(\omega)=K_0g,\;\;\;\omega>0
\end{equation}
where $f_{c,\beta}:(0,\infty)\rightarrow \Real$ is defined by
\begin{equation}\label{df}
f_{c,\beta}(\omega)=\frac{1}{\rho_c(\omega)}\Big[
\frac{\omega}{2}-\frac{2\beta}{\omega}\Big].
\end{equation}

We can thus write an explicit expression for the velocity vs.
synaptic-coupling strength curve, defined by (\ref{defsig}), as
follows \begin{equation}\label{repsig0}
\Sigma=\{(\omega,f_{c,\beta}(\omega))\;|\; \omega>0\}.
\end{equation}

In the following lemma we collect some properties of the functions
$f_{c,\beta}(\omega)$, which are obtained by elementary calculus:

\begin{lemma}\label{prf} We have
\begin{equation}\label{lz}\lim_{\omega\rightarrow
+\infty}{f_{c,\beta}(\omega)}=+\infty,
\end{equation}

\noindent and

\noindent (i) When $\beta<0$, $f_{c,\beta}$ is positive and convex
on $(0,\infty)$, and

\begin{equation}\label{li}\lim_{\omega\rightarrow 0+}{f_{c,\beta}(\omega)}=+\infty.\end{equation}

\noindent (ii) When $\beta>0$, $f_{c,\beta}$ is increasing on
$(0,\infty)$, and
\begin{equation}\label{li1}\lim_{\omega\rightarrow 0+}{f_{c,\beta}(\omega)}=-\infty.\end{equation}

\noindent (iii) When $\beta=0$, $f_{c,\beta}$ is increasing on
$(0,\infty)$, and
\begin{equation}\label{li2}\lim_{\omega\rightarrow 0+}{f_{c,\beta}(\omega)}=0\end{equation}
\end{lemma}

Due to (\ref{prf}) and part (i) of lemma \ref{prf}, we can define,
for $\beta<0$,
\begin{equation}\label{defom}
\Omega(c,\beta)=\min_{\omega>0}{f_{c,\beta}}(\omega).
\end{equation}

From lemma \ref{prf} we conclude that

\begin{lemma}\label{solutions}
\noindent (i) When $\beta<0$ the equation
\begin{equation}\label{eqy}
f_{c,\beta}(\omega)=y,\;\;\;\omega>0
\end{equation}
has exactly two solutions when $y>\Omega(c,\beta)$, which we will
denote by
$$\underline{\omega}_{c,\beta}(y)<\overline{\omega}_{c,\beta}(y),$$
a unique solution when $y=\Omega(c,\beta)$, and no solutions when
$y<\Omega(c,\beta)$.

\noindent (ii) When $\beta\geq 0$ the equation (\ref{eqy}) has a
unique solution for all $y\in\Real$ in the case $\beta>0$ and for
all $y>0$ in the case $\beta=0$, which we will denote by
$\omega_{c,\beta}(y)$.
\end{lemma}

An elementary asymptotic analysis of the equation (\ref{eqy})
yields

\begin{lemma}\label{as}
(I) When $\beta<0$ we have
\begin{equation}\label{asv1}
\underline{\omega}_{c,\beta}(y)=2|\beta|e^{\frac{c}{2}}\frac{1}{y}+O\Big(\frac{1}{y^3}\Big)\;\;\;as\;\;y\rightarrow\infty,
\end{equation}
and when $c>0$
\begin{equation}\label{asv2}
\overline{\omega}_{c,\beta}(y)=2\sqrt{\frac{\pi}{e^{\frac{c}{2}}-1}}\sqrt{y}+O(1)\;\;\;as\;\;y\rightarrow\infty,
\end{equation}
while when $c=0$
\begin{equation}\label{asv22}
\overline{\omega}_{0,\beta}(y)=2
y+O\Big(\frac{1}{y}\Big)\;\;\;as\;\;y\rightarrow\infty.
\end{equation}

\noindent (II) When $\beta>0$: As $y\rightarrow +\infty$
$\omega_{c,\beta}(y)$ has the same asymptotic behavior as in
(\ref{asv2}),(\ref{asv22}) in the cases $c>0$, $c=0$,
respectively. As $y\rightarrow -\infty$ we have
\begin{equation}\label{asv33}
\omega_{c,\beta}(y)=-4\beta
e^{\frac{c}{2}}\frac{1}{y}+O\Big(\frac{1}{y^3}\Big)\;\;\;as\;\;y\rightarrow
-\infty.
\end{equation}

\noindent (III) When $\beta=0$: As $y\rightarrow +\infty$
$\omega_{c,\beta}(y)$ has the same asymptotic behavior as in
(\ref{asv2}),(\ref{asv22}) in the cases $c>0$, $c=0$,
respectively. As $y\rightarrow 0$, we have
\begin{equation}\label{asv44}
\overline{\omega}(y)=o(1)\;\;\;as\;\;y\rightarrow 0.
\end{equation}
\end{lemma}

Summarizing our results on the case of uniform coupling, we have

\begin{theorem}\label{unij} When $K(x)\equiv K_0$:

\noindent (I) In the excitable case $\beta<0$:

\noindent (i) If $g>\frac{\Omega(c,\beta)}{K_0}$ there exist two
rotating waves with velocities given by
$\underline{\omega}_{c,\beta}(K_0g)$ and
$\overline{\omega}_{c,\beta}(K_0g)$, and we have, for the slow
wave
\begin{equation}\label{asv10}
\underline{\omega}_{c,\beta}(K_0g)=\frac{2|\beta|e^{\frac{c}{2}}}{K_0}\frac{1}{g}+O\Big(\frac{1}{g^3}\Big)\;\;\;as\;\;g\rightarrow\infty,
\end{equation}
for the fast wave when $c>0$:
\begin{equation}\label{asv20}
\overline{\omega}_{c,\beta}(K_0g)=2\sqrt{\frac{\pi
K_0}{e^{\frac{c}{2}}-1}}\sqrt{g}+O(1)\;\;\;as\;\;g\rightarrow\infty.
\end{equation}
while for the fast wave when $c=0$
\begin{equation}\label{asv220}
\overline{\omega}_{0,\beta}(K_0g)=2K_0
g+O\Big(\frac{1}{g}\Big)\;\;\;as\;\;g\rightarrow\infty.
\end{equation}
\noindent (ii) If $g=\frac{\Omega(c,\beta)}{K_0}$ there exists a
unique rotating wave with velocity
\begin{equation}\label{asv3}
\omega=\underline{\omega}_{c,\beta}(K_0g)=\overline{\omega}_{c,\beta}(
K_0g).
\end{equation}
\noindent (iii) If $g<\frac{\Omega(c,\beta)}{K_0}$ there exist no
rotating waves.

\noindent (II) In the oscillatory case $\beta> 0$, there exists a
unique rotating wave for any $g\in \Real$, whose velocity is given
by $\omega_{c,\beta}(K_0g)$, and for $g\rightarrow +\infty$ it has
the same asymptotics as in (\ref{asv20}),(\ref{asv220}) in the
cases $c>0$, $c=0$, respectively, while for $g\rightarrow -\infty$
\begin{equation}\label{asv330}
\overline{\omega}(K_0g)=-\frac{4\beta
e^{\frac{c}{2}}}{K_0}\frac{1}{g}+O\Big(\frac{1}{g^3}\Big)\;\;\;as\;\;g\rightarrow
-\infty.
\end{equation}

\noindent (III) In the boundary case $\beta=0$:

\noindent (i) For $g>0$ there exists a unique rotating wave, whose
velocity is given by $\omega_{c,\beta}(K_0g)$, and for
$g\rightarrow\infty$ it has the same asymptotics as in
(\ref{asv20}),(\ref{asv220}) in the cases $c>0$, $c=0$,
respectively, while for $g\rightarrow 0$
\begin{equation}\label{asv440}
\overline{\omega}(K_0g)=o(1)\;\;\;as\;\;g\rightarrow 0.
\end{equation}

\noindent (ii) If $g\leq 0$ there exist no rotating waves.
\end{theorem}

\vspace{0.4cm} We now observe that in the special case $c=0$ (the
model introduced in \cite{iz}) we can obtain more explicit
expressions. Using (\ref{rho0}),(\ref{df}) we have
$$f_{0,\beta}(\omega)=\frac{\omega}{2}-\frac{2\beta}{\omega}.$$
The minimum in (\ref{defom}) can now be computed explicitly, and we obtain, when
$\beta<0$,
$$\Omega(0,\beta)=2\sqrt{|\beta|}.$$
We can also solve (\ref{eqc1}) explicitly, and obtain the
velocities of the rotating waves. When $\beta<0$,
$g>\frac{2\sqrt{|\beta|}}{K_0}$
$$\underline{\omega}_{0,\beta}(g)=K_0g-\sqrt{(K_0g)^2+4\beta},\;\;\;\overline{\omega}_{0,\beta}(g)=K_0g+\sqrt{(K_0g)^2+4\beta}.$$
When $\beta\geq 0$
$$\omega_{0,\beta}(g)=\sqrt{(K_0g)^2+4\beta}+K_0g.$$

\begin{figure}
\centering
    \includegraphics[height=7cm,width=7cm, angle=0]{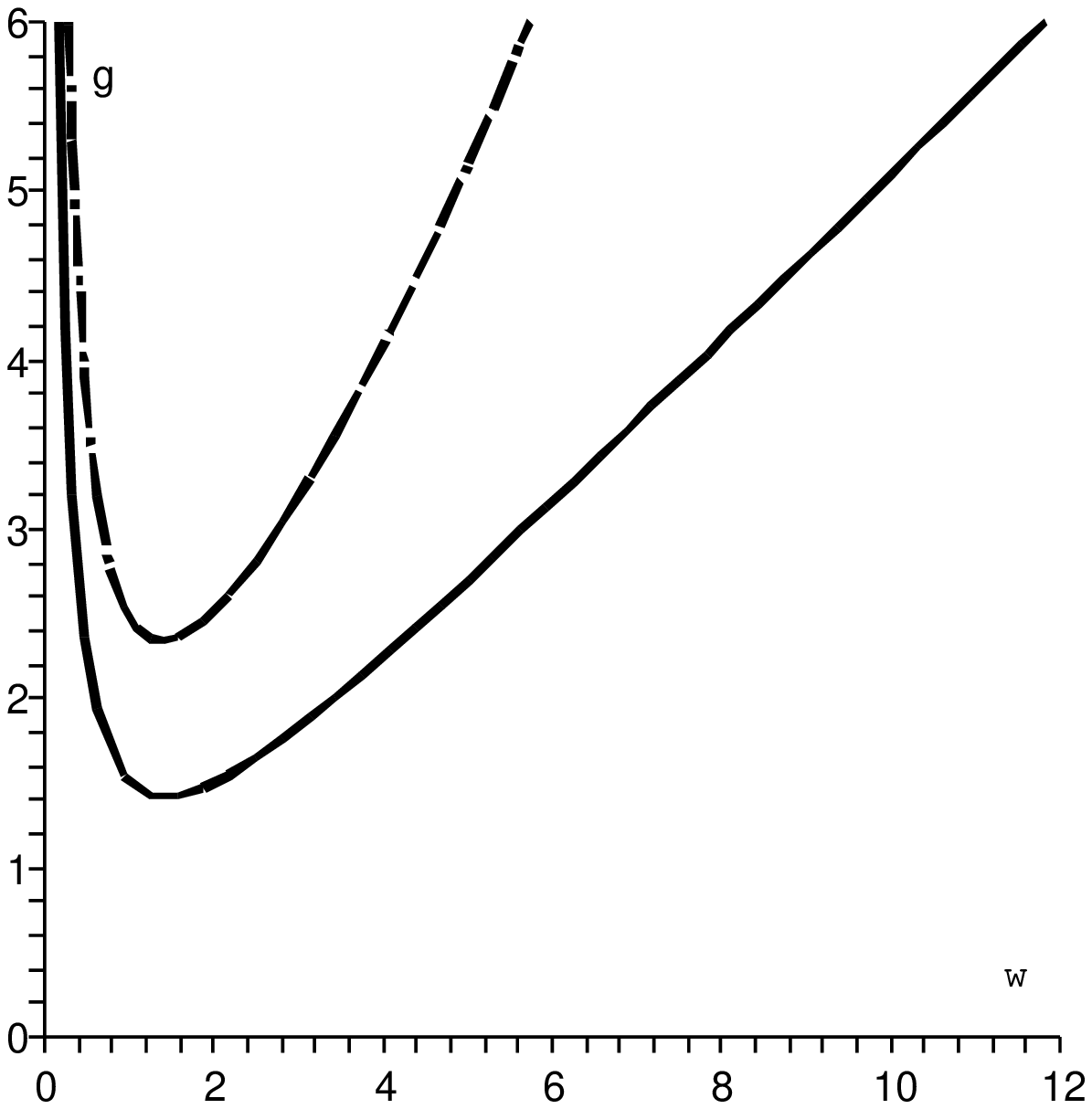}
    \caption{The $\omega$-$g$ curves ($\Sigma$) when $K(x)\equiv 1$, $\beta=-0.5$, for the cases $c=0$, $c=1$ (dashed line).}
    \label{syn1}
\end{figure}

\begin{figure}
\centering
    \includegraphics[height=7cm,width=7cm, angle=0]{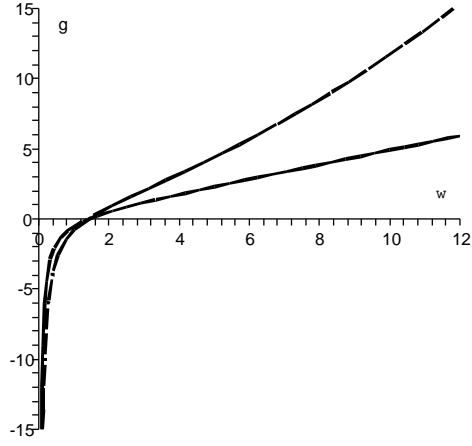}
    \caption{The $\omega$-$g$ curves ($\Sigma$) when $K(x)\equiv 1$, $\beta=0.5$, for the cases $c=0$, $c=1$ (dashed line).}
    \label{syn2}
\end{figure}

\vspace{0.4cm} Figures \ref{syn1},\ref{syn2} show the curves
$\Sigma$ curves for the rotating waves when $K(x)\equiv 1$, in an
excitable ($\beta=-0.5$) and an oscillatory ($\beta=0.5$) case,
for $c=0$ and $c=1$.

\section{Rotating waves on a ring in the general case}
\label{general}

We now return to the case when $K\in L^{\infty}(\Real)$ is a
function satisfying (\ref{jpos}),(\ref{jper}) (these will be
standing assumptions throughout this section and the next one),
and prove that several of the results about rotating waves
obtained above for the special case of uniform coupling remain
valid, though the proofs are necessarily less direct.

Our main theorem generalizes the representation (\ref{repsig0})
for $\Sigma$ which was obtained in the case of uniform coupling.
The function $f_{c,\beta}(\omega)$ will be replaced by a function
$g(\omega)$, for which, unlike in the uniform-coupling case, we
cannot obtain an explicit expression, but we shall be able to
derive some key qualitative properties, which are similar to those
of $f_{c,\beta}$.

\begin{theorem}\label{curve}
The synaptic-coupling strength ($g$) vs. velocity ($\omega$) curve
$\Sigma$, defined by (\ref{defsig}) can be represented as the
graph of a function:
\begin{equation}\label{repsig}
\Sigma=\{(\omega,g(\omega)) \;|\; \omega\in (0,\infty)\},
\end{equation}
where $g:(0,\infty)\rightarrow \Real$ is a continuous function and

\noindent we have
\begin{equation}\label{g00}
\lim_{\omega\rightarrow +\infty}g(\omega)=+\infty,
\end{equation}

\noindent and

\noindent (i) When $\omega^2\geq 4\beta$ we have the inequalities:
\begin{equation}\label{pp1}\frac{1}{\overline{K}}f_{c,\beta}(\omega)\leq g(\omega)\leq
\frac{1}{\underline{K}}f_{c,\beta}(\omega),\end{equation}
\begin{equation}\label{pp2}
\frac{2}{\|K\|_{L^1[0,2\pi]}}\Big(\frac{\omega^2}{4}-\beta
\Big)\Big(e^{\frac{c}{2}+\frac{2\pi}{\omega}}-1\Big)e^{-\frac{2\pi}{\omega}}
\leq g(\omega) \leq \frac{2}{\|K\|_{L^1[0,2\pi]}}\Big(
\frac{\omega^2}{4}-\beta\Big)\Big(e^{\frac{c}{2}+\frac{2\pi}{\omega}}-
1\Big).
\end{equation}

\noindent (ii) When $\omega^2\leq 4\beta$ we have the
inequalities:
\begin{equation}\label{pp3}\frac{1}{\underline{K}}f_{c,\beta}(\omega)\leq g(\omega)\leq
\frac{1}{\overline{K}}f_{c,\beta}(\omega),
\end{equation}
\begin{equation}\label{pp4}
\frac{2}{\|K\|_{L^1[0,2\pi]}}\Big(
\frac{\omega^2}{4}-\beta\Big)\Big(e^{\frac{c}{2}+\frac{2\pi}{\omega}}-
1\Big) \leq g(\omega) \leq \frac{2}{\|K\|_{L^1[0,2\pi]}}\Big(
\frac{\omega^2}{4}-\beta\Big)\Big(e^{\frac{c}{2}+\frac{2\pi}{\omega}}-1\Big)e^{-\frac{2\pi}{\omega}}.
\end{equation}

\noindent and

\noindent (I) In the excitable case $\beta<0$
$$\lim_{\omega\rightarrow 0+}g(\omega)=+\infty.$$

\noindent (II) In the oscillatory case $\beta>0$
$$\lim_{\omega\rightarrow 0+}g(\omega)=-\infty.$$

\noindent (III) In the borderline case $\beta=0$
$$\lim_{\omega\rightarrow 0+}g(\omega)=0.$$

\end{theorem}

Sketching the graphs of $g(\omega)$ as described by theorem
\ref{curve} in the cases $\beta>0$, $\beta<0$, $\beta=0$, shows
that theorem \ref{curve} implies theorem \ref{sum}, and we note
that in the case $\beta<0$ we have
\begin{equation}\label{gcrit}
g_{crit}=\min_{\omega>0}{g(\omega)}.
\end{equation}

\begin{figure}
\centering
    \includegraphics[height=7cm,width=7cm, angle=0]{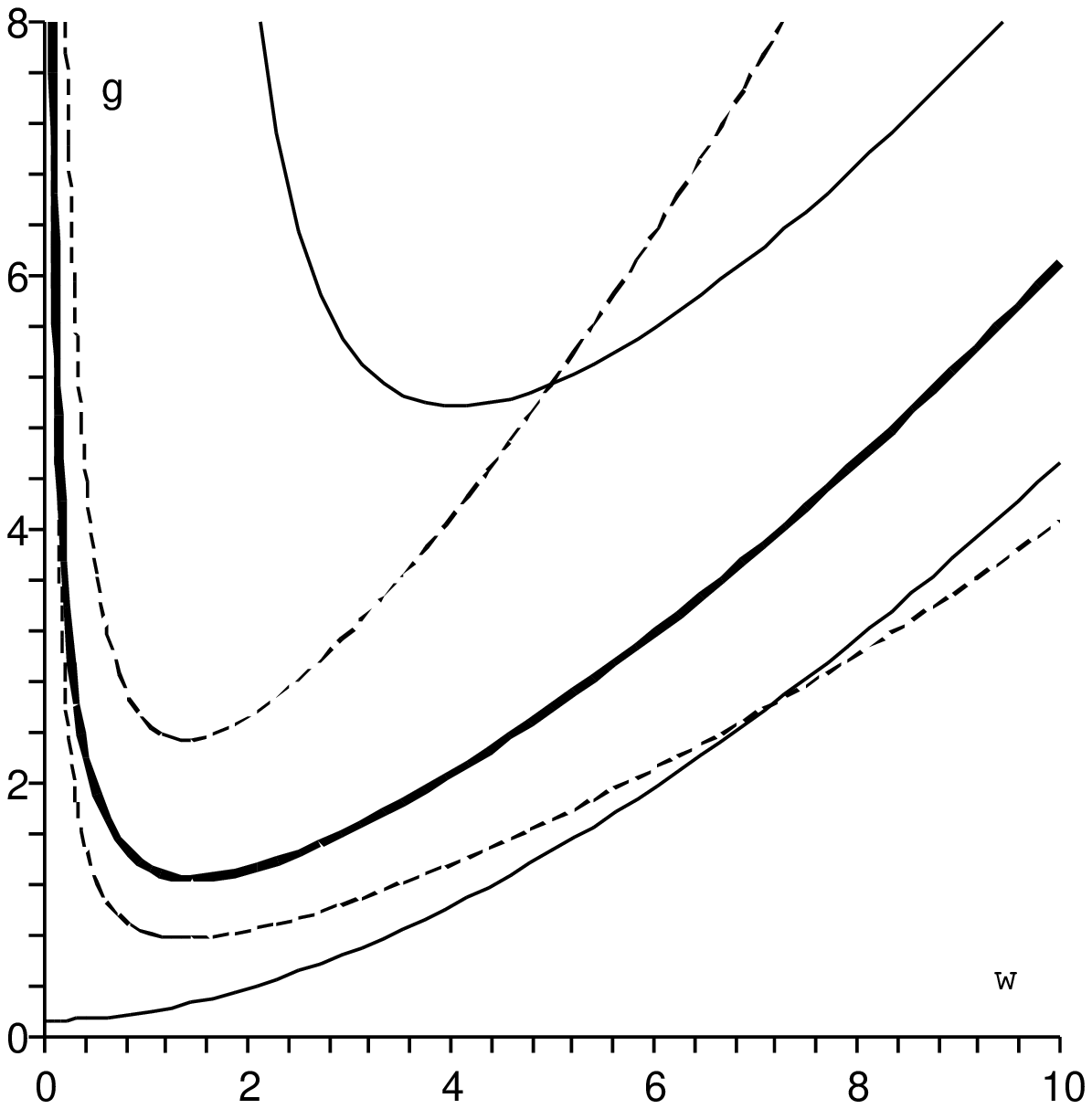}
    \caption{Graph of the function $g(\omega)$ for $K(x)=2+\cos(x), \beta=-0.5, c=1$, obtained numerically (bold curve),
    together with the graphs of the bounds given by (\ref{pp1}) (dashed curves) and by (\ref{pp2})
    (lighter
    curves)}.
    \label{il1}
\end{figure}

\begin{figure}
\centering
    \includegraphics[height=7cm,width=7cm, angle=0]{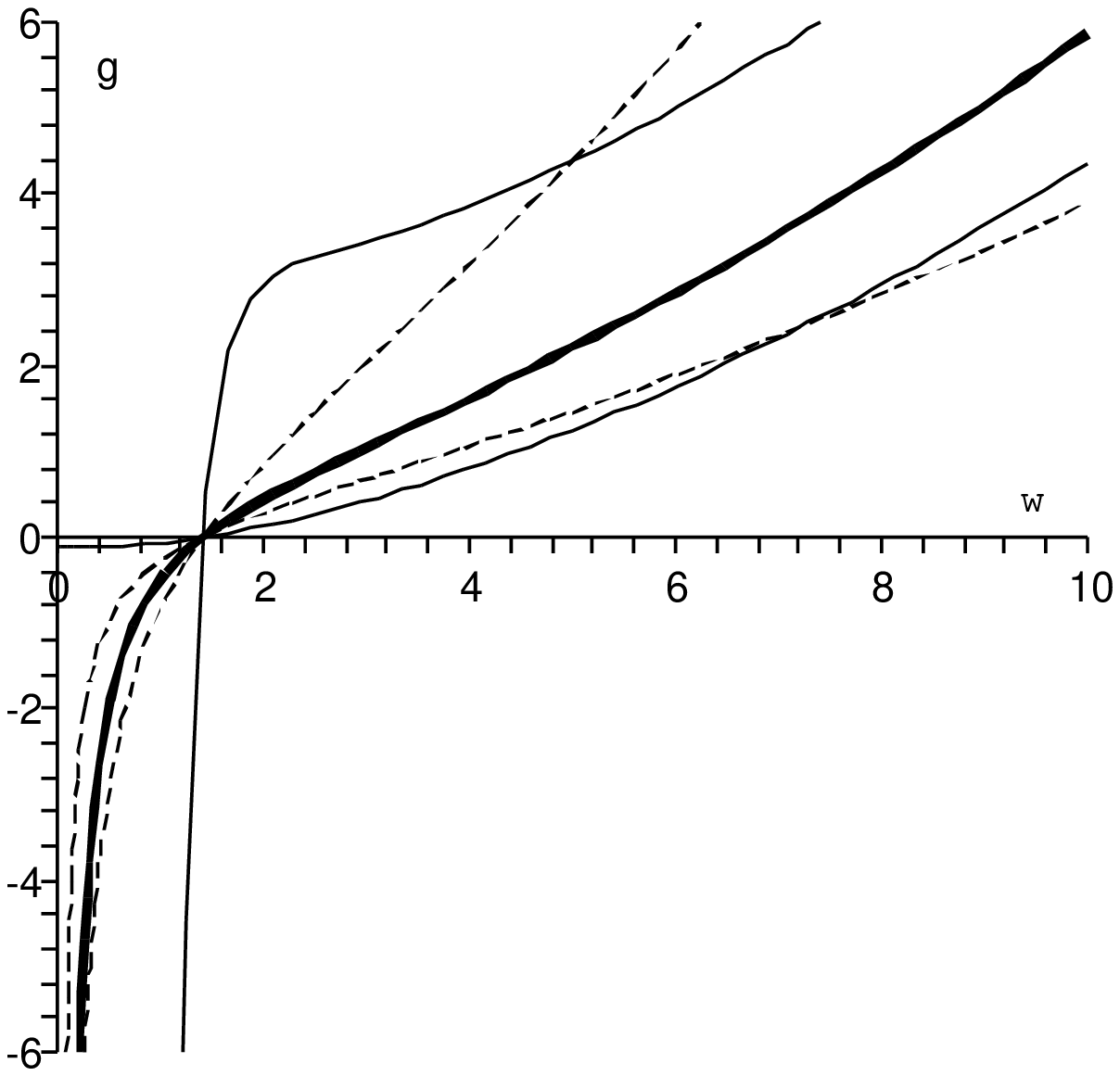}
    \caption{Graph of the function $g(\omega)$ for $K(x)=2+\cos(x), \beta=0.5, c=1$, obtained numerically (bold curve),
    together with the graphs of the bounds given by (\ref{pp1}), (\ref{pp3}) (dashed curves) and by
    (\ref{pp2}),(\ref{pp4})
    (lighter
    curves)}.
    \label{il2}
\end{figure}

Figure \ref{il1} illustrates the contents of theorem \ref{curve}
in the excitable case. In this example we take $K(x)=2+\cos(x),
\beta=-0.5, c=1$. The bold curve shows the graph of the function
$g(\omega)$ (or in other words the curve $\Sigma$), obtained by
laborious numerical computation: we numerically solved the
equation (\ref{tt}) for $g$, with the parameter $\omega$ taking
values in the interval $(0,10)$ - using the bisection method. Note
that each evaluation of the function $\Psi$ requires numerically
solving the initial-value problem (\ref{tww}), (\ref{ip}). The
dashed curves in the figure represent the lower and upper bounds
on $g(\omega)$ given by (\ref{pp1}), while the lighter curves
represent the lower and upper bounds given by (\ref{pp2}).

Figure \ref{il2} illustrates the contents of theorem \ref{curve}
in the oscillatory case. We take $K(x)=2+\cos(x), \beta=-0.5,
c=1$. The bold curve shows the graph of the function $g(\omega)$,
obtained numerically, The dashed curves in the figure represent
the lower and upper bounds on $g(\omega)$ given by
(\ref{pp1}),(\ref{pp3}) while the lighter curves represent the
lower and upper bounds given by (\ref{pp2}),(\ref{pp4}).

\vspace{0.4cm}

We prove theorem \ref{curve} by using the following lemmas, whose
proofs will follow.

We define the functions
$$\underline{g}(\omega)=
\frac{1}{\overline{R}_{\omega}}\Big(\frac{\omega^2}{4}-\beta\Big),\;\;\;
\overline{g}(\omega)=
\frac{1}{\underline{R}_{\omega}}\Big(\frac{\omega^2}{4}-\beta\Big).$$

\begin{lemma}\label{ineqq}
\noindent (i) If $\omega^2\geq 4\beta$ then
\begin{equation}\label{p1}
0\leq \frac{1}{\overline{K}}f_{c,\beta}(\omega)\leq
\underline{g}(\omega)\leq \overline{g}(\omega)\leq
\frac{1}{\underline{K}}f_{c,\beta}(\omega),
\end{equation}
\begin{equation}\label{p2}
\frac{2}{\|K\|_{L^1[0,2\pi]}}\Big(\frac{\omega^2}{4}-\beta
\Big)\Big(e^{\frac{c}{2}+\frac{2\pi}{\omega}}-1\Big)e^{-\frac{2\pi}{\omega}}
\leq \underline{g}(\omega)\leq \overline{g}(\omega) \leq
\frac{2}{\|K\|_{L^1[0,2\pi]}}\Big(
\frac{\omega^2}{4}-\beta\Big)\Big(e^{\frac{c}{2}+\frac{2\pi}{\omega}}-
1\Big).
\end{equation}

\noindent (ii) If $\omega^2\leq 4\beta$ then
\begin{equation}\label{p3}
\frac{1}{\underline{K}}f_{c,\beta}(\omega)\leq
\overline{g}(\omega)\leq \underline{g}(\omega)\leq
\frac{1}{\overline{K}}f_{c,\beta}(\omega)\leq 0,
\end{equation}
\begin{equation}\label{p4}
\frac{2}{\|K\|_{L^1[0,2\pi]}}\Big(
\frac{\omega^2}{4}-\beta\Big)\Big(e^{\frac{c}{2}+\frac{2\pi}{\omega}}-
1\Big)\leq \underline{g}(\omega) \leq
\frac{2}{\|K\|_{L^1[0,2\pi]}}\Big(\frac{\omega^2}{4}-\beta
\Big)\Big(e^{\frac{c}{2}+\frac{2\pi}{\omega}}-1\Big)e^{-\frac{2\pi}{\omega}}.
\end{equation}
\end{lemma}

\begin{lemma}
\label{noex} If either

\noindent (i) $g\geq 0$ and $g<\underline{g}(\omega)$

\noindent or

\noindent (ii) $g\leq 0$ and $g<\overline{g}(\omega)$

\noindent then
\begin{equation}\label{sm}
\Psi(\omega,g)<1.
\end{equation}
\end{lemma}

\begin{lemma}\label{glar}
If either

\noindent (i) $g\geq 0$ and $g>\overline{g}(\omega)$.

\noindent or

\noindent (ii) $g\leq 0$ and $g>\underline{g}(\omega)$.

\noindent then
\begin{equation}\label{conc1}
\Psi(\omega,g)>1.
\end{equation}
\end{lemma}

\begin{lemma}\label{unil}
Fixing $\omega>0$, there is {\bf{at most}} one value of $g$ for
which (\ref{tt}) holds.
\end{lemma}

Using the above lemmas, we can give the

\noindent {\sc{proof of theorem \ref{curve}:}}  First we fix
$\omega> 0$ satisfying $\omega^2\geq 4\beta$. Then by part (i) of
lemma \ref{ineqq} we have
$0\leq\underline{g}(\omega)\leq\overline{g}(\omega)$, hence by
part (i) of lemma \ref{noex}, part (i) of lemma \ref{glar} and the
intermediate-value theorem, there exists $g$ satisfying
$\underline{g}(\omega) \leq g \leq \overline{g}(\omega)$ with
$\Psi(\omega,g)=1$. By lemma \ref{unil} this $g$ is unique, so we
may denote it by $g(\omega)$.

Now fix $\omega>0$ satisfying $\omega^2\leq 4\beta$. Then by part
(ii) of lemma \ref{ineqq} we have
$\overline{g}(\omega)\leq\underline{g}(\omega)\leq 0$, hence by
part (ii) of lemma \ref{noex}, part (ii) of lemma \ref{glar} and
the intermediate-value theorem, there exists $g$ satisfying
$\overline{g}(\omega) \leq g\leq \underline{g}(\omega)$ with
$\Psi(\omega,g)=1$. By lemma (\ref{unil}) this $g$ is unique, so
we may denote it by $g(\omega)$.

We have thus proven the existence of $g(\omega)$ satisfying
(\ref{repsig}), and
\begin{equation}\label{inn1}
\omega^2\geq 4\beta \;\Rightarrow 0\leq\underline{g}(\omega)\leq
g(\omega)\leq \overline{g}(\omega).
\end{equation}
\begin{equation}\label{inn2}
\omega^2\leq 4\beta \;\Rightarrow \overline{g}(\omega)\leq
g(\omega)\leq \underline{g}(\omega)\leq 0.
\end{equation}

Parts (i),(ii) of theorem \ref{curve} follow from
(\ref{inn1}),(\ref{inn2}) and lemma \ref{ineqq}.

(\ref{g00}) follows from part (i) of the theorem, and (\ref{lz}).

Parts (I)-(III) of theorem \ref{curve} follow from (\ref{pp1}) and
(\ref{pp3}), along with (\ref{li}), (\ref{li1}) and (\ref{li2}).

To conclude the proof of theorem \ref{curve} we only need to show
that $g(\omega)$ is continuous, but this follows at once from
(\ref{repsig}), since by continuity of $\Psi$, the set $\Sigma$ is
closed, and this set is the graph of $g(\omega)$.

\vspace{0.4cm} We now turn to the proofs of lemmas
\ref{ineqq}-\ref{unil}.

Lemma \ref{ineqq} is an immediate consequence of the definitions
of $\underline{g}(\omega)$,$\overline{g}(\omega)$ and of the
inequalities of lemmas \ref{ulb} and \ref{ulb1}.

We now start working towards proving lemma \ref{noex}.

\begin{lemma}
\label{ell}
If either

\noindent (i) $g\geq 0$ and
\begin{equation}\label{cl}
\beta+g\overline{R}_{\omega}<0
\end{equation}
or

\noindent (ii) $g\leq 0$ and
\begin{equation}\label{cl1}
\beta+g\underline{R}_{\omega}<0
\end{equation}
then
$$\Psi(\omega,g)<\frac{2}{3}.$$
\end{lemma}

\noindent {\sc{proof:}}  (i) We will show that when
(\ref{cl}) holds we have
\begin{equation}\label{cons1}
\phi_{\omega,g}(z)<2\pi\;\;\;\forall z\geq 0,
\end{equation}
which, together with (\ref{aps}),  implies the result of our
lemma. To prove our claim we note that, using
(\ref{tww}),(\ref{hz}),(\ref{wnn}),(\ref{cl}), and the assumption
that $g\geq 0$
\begin{eqnarray}\label{kee}
\phi_{\omega,g}(z)=2\pi \;&\Rightarrow\;&
\phi_{\omega,g}'(z)=\frac{2}{\omega}[\beta+g
R_{\omega}(z)]\nonumber\\&\leq& \frac{2}{\omega}[\beta+
g\overline{R}_{\omega}] < 0.
\end{eqnarray}
We now show that (\ref{kee}) implies (\ref{cons1}). If
(\ref{cons1}) fails to hold, then we set
$$z_0=\min \;\{z\geq 0 \;|\; \phi_{\omega,g}(z)= 2\pi\}.$$
This number is well-defined by continuity and by the fact that
$\phi_{\omega,g}(0)=\pi$, which implies also that $z_0>0$. By
(\ref{kee}) we have $\phi_{\omega,g}'(z_0)<0$, but this implies
that $\phi_{\omega,g}(z)$ is decreasing in a neighborhood of
$z_0$, and in particular that there exist $z\in (0,z_0)$
satisfying $\phi_{\omega,g}(z)=2\pi.$ But this contradicts the
definition of $z_0$, and this contradiction proves (\ref{cons1}).

\noindent (ii) In the case that $g\leq 0$ and (\ref{cl1}) holds,
we replace (\ref{kee}) with
\begin{eqnarray*}
\phi_{\omega,g}(z)=2\pi \;&\Rightarrow\;&
\phi_{\omega,g}'(z)=\frac{2}{\omega}[\beta+g
R_{\omega}(z)]\nonumber\\&\leq& \frac{2}{\omega}[\beta+
g\underline{R}_{\omega}] < 0.
\end{eqnarray*}
so that the proof proceeds as before.

\vspace{0.4cm}

\noindent {\sc{proof of lemma \ref{noex}:}} (i) We assume that $g
\geq 0$ and $g<\underline{g}(\omega)$, {\it{i.e.}}
\begin{equation}\label{asi}
g<\frac{1}{\overline{R}_{\omega}}\Big(\frac{\omega^2}{4}-\beta\Big)
\end{equation}
and prove (\ref{sm}).

By (\ref{aps}), (\ref{sm}) is equivalent to
\begin{equation}
\label{din0} \phi_{\omega,g}(2\pi)<3\pi.
\end{equation}
We note that, by part (i) of lemma \ref{ell}, we already have
(\ref{sm}) when (\ref{cl}) holds, hence we may assume
\begin{equation}
\label{aaa} \overline{\mu}=\beta+g\overline{R}_{\omega}\geq 0.
\end{equation}
Using (\ref{tww}) and the assumption that $g\geq 0$, we have
\begin{eqnarray}\label{mmm}
\phi_{\omega,g}'(z)&=&\frac{1}{\omega} [h(\phi_{\omega,g}(z))+ g
R_{\omega}(z) w(\phi_{\omega,g}(z))] \nonumber\\&\leq&
\frac{1}{\omega} \Big[1-\cos(\phi_{\omega,g}(z))+
\Big(\beta+g\overline{R}_{\omega}\Big)
(1+\cos(\phi_{\omega,g}(z)))\Big]\nonumber\\&=&
\frac{1}{\omega}[(\overline{\mu}
+1)+(\overline{\mu}-1)\cos(\phi_{\omega,g}(z))].
\end{eqnarray}
which implies (note that the integral below is well-defined
because $\overline{\mu}\geq 0$ - if $\overline{\mu}=0$ the
integral is infinite)
$$\int_{0}^{2\pi}{\frac{\phi_{\omega,g}'(z)dz}{(\overline{\mu}
+1)+(\overline{\mu}-1)\cos(\phi_{\omega,g}(z))}}\leq
\frac{2\pi}{\omega} .$$ Making the change of variables
$\varphi=\phi_{\omega,g}(z)$ we obtain
\begin{equation}\label{ha0}
\int_{\phi_{\omega,g}(0)}^{\phi_{\omega,g}(2\pi)}
{\frac{d\varphi}{(\overline{\mu}
+1)+(\overline{\mu}-1)\cos(\varphi)}}\leq \frac{2\pi}{\omega}.
\end{equation}
If we assume, by way of contradiction, that (\ref{din0}) does
{\it{not}} hold, {\it{i.e.}} that $\phi_{\omega,g}(2\pi)\geq
3\pi$, then, using (\ref{form}),
$$\int_{\phi_{\omega,g}(0)}^{\phi_{\omega,g}(2\pi)}{\frac{d\varphi}{(\overline{\mu}
+1)+(\overline{\mu}-1)\cos(\varphi)}}\geq
\int_{\pi}^{3\pi}{\frac{d\varphi}{(\overline{\mu}
+1)+(\overline{\mu}-1)\cos(\varphi)}}=\frac{\pi}{\sqrt{\overline{\mu}}},$$
so together with (\ref{ha0}) we obtain
$$\frac{1}{\sqrt{\overline{\mu}}}\leq \frac{2}{\omega},$$
which is equivalent to
$$g\geq \frac{1}{\overline{R}_{\omega}}\Big(\frac{\omega^2}{4}-\beta\Big),$$
which contradicts (\ref{asi}). This contradiction proves
(\ref{din0}), concluding the proof that (i) implies (\ref{sm}).

\noindent (ii) We assume that $g\leq 0$ and
$g<\overline{g}(\omega)$, {\it{i.e.}}
\begin{equation}\label{asi1}
g<\frac{1}{\underline{R}_{\omega}}\Big(\frac{\omega^2}{4}-\beta\Big)
\end{equation}
and prove (\ref{din0}).

We note that, by part (ii) of lemma \ref{ell}, we already have
(\ref{sm}) when (\ref{cl1}) holds, hence we may assume
$$\underline{\mu}=\beta+g\underline{R}_{\omega}\geq 0.$$

We now replace (\ref{mmm}) by
\begin{eqnarray*}
\phi_{\omega,g}'(z)&=&\frac{1}{\omega} [h(\phi_{\omega,g}(z))+ g
R_{\omega}(z) w(\phi_{\omega,g}(z))] \nonumber\\&\leq&
\frac{1}{\omega} \Big[1-\cos(\phi_{\omega,g}(z))+
\Big(\beta+g\underline{R}_{\omega}\Big)
(1+\cos(\phi_{\omega,g}(z)))\Big]\nonumber\\&=&
\frac{1}{\omega}[(\underline{\mu}
+1)+(\underline{\mu}-1)\cos(\phi_{\omega,g}(z))].
\end{eqnarray*}
The proof then proceeds as before.

\vspace{0.4cm} \noindent {\sc{proof of lemma \ref{glar}:}} (i) We
assume $g\geq 0$ and $g>\overline{g}(\omega)$, {\it{i.e.}},
\begin{equation}\label{kein}
g>\frac{1}{\underline{R}_{\omega}}\Big(\frac{\omega^2}{4}-\beta\Big)
\end{equation} and prove that $\Psi(\omega,g)>1$.
By (\ref{aps}), our claim is equivalent to
\begin{equation}
\label{din} \phi_{\omega,g}(2\pi)>3\pi.
\end{equation}
We define
$$\underline{\mu}=\beta+g\underline{R}_{\omega},$$
and we note that (\ref{kein})
is equivalent to
\begin{equation}
\label{eg2}\underline{\mu}>\frac{\omega^2}{4}.
\end{equation}
Using (\ref{tww}) and $g\geq 0$ we have
\begin{eqnarray}\label{lll}
\phi_{\omega,g}'(z)&=&\frac{1}{\omega} [h(\phi_{\omega,g}(z))+ g
R_{\omega}(z) w(\phi_{\omega,g}(z))] \nonumber\\&\geq&
\frac{1}{\omega}\Big[1-\cos(\phi_{\omega,g}(x))+
\Big(\beta+g\underline{R}_{\omega}\Big)
(1+\cos(\phi_{\omega,g}(z)))\Big]\nonumber\\&=& \frac{1}{\omega}
[(\underline{\mu}
+1)+(\underline{\mu}-1)\cos(\phi_{\omega,g}(z))],
\end{eqnarray}
which implies
$$\int_{\pi}^{3\pi}{\frac{\phi_{\omega,g}'(z)dz}{(\underline{\mu}
+1)+(\underline{\mu}-1)\cos(\phi_{\omega,g}(z))}}\geq
\frac{2\pi}{\omega} .$$ Making the change of variables
$\varphi=\phi_{\omega,g}(z)$, we obtain
\begin{equation}\label{ha}\int_{\phi_{\omega,g}(0)}^{\phi_{\omega,g}(2\pi)}
{\frac{d\varphi}{(\underline{\mu}
+1)+(\underline{\mu}-1)\cos(\varphi)}}\geq \frac{2\pi}{\omega}.
\end{equation}
If we assume, by way of contradiction, that (\ref{din}) does
{\it{not}} hold, {\it{i.e.}} that $\phi_{\omega,g}(2\pi)\leq
3\pi$, then, using (\ref{form}),
$$\int_{\phi_{\omega,g}(0)}^{\phi_{\omega,g}(2\pi)}{\frac{d\varphi}{(\underline{\mu}
+1)+(\underline{\mu}-1)\cos(\varphi)}}\leq
\int_{\pi}^{3\pi}{\frac{d\varphi}{(\underline{\mu}
+1)+(\underline{\mu}-1)\cos(\varphi)}}=\frac{\pi}{\sqrt{\underline{\mu}}},$$
so together with (\ref{ha}) we obtain
$$\frac{1}{\sqrt{\underline{\mu}}}\geq \frac{2}{\omega}.$$
This contradicts (\ref{eg2}), and this contradiction implies that
(\ref{din}) holds, completing our proof that (i) implies
(\ref{conc1}).

\noindent (ii) We assume $g\leq 0$ and $g>\underline{g}(\omega)$,
{\it{i.e.}},
\begin{equation}\label{keinn}
g>\frac{1}{\overline{R}_{\omega}}\Big(\frac{\omega^2}{4}-\beta\Big)
\end{equation} and prove (\ref{conc1}).
We define
$$\overline{\mu}=\beta+g\overline{R}_{\omega},$$
and we note that (\ref{keinn}) is equivalent to
$$\overline{\mu}>\frac{\omega^2}{4}.$$

Using (\ref{tww}) and $g\leq 0$ we have
\begin{eqnarray*}
\phi_{\omega,g}'(z)&=&\frac{1}{\omega} [h(\phi_{\omega,g}(z))+ g
R_{\omega}(z) w(\phi_{\omega,g}(z))] \nonumber\\&\geq&
\frac{1}{\omega}\Big[1-\cos(\phi_{\omega,g}(x))+
\Big(\beta+g\overline{R}_{\omega}\Big)
(1+\cos(\phi_{\omega,g}(z)))\Big]\nonumber\\&=& \frac{1}{\omega}
[(\overline{\mu} +1)+(\overline{\mu}-1)\cos(\phi_{\omega,g}(z))],
\end{eqnarray*}
and the rest of the proof proceeds as in case (i) above.

\vspace{0.4cm}

To prove lemma \ref{unil} we shall prove the following more
general lemma - from which lemma \ref{unil} follows by setting
$R(z)=R_{\omega}(z)$, and which we shall also have occasion to use
once more later.

\begin{lemma}\label{unig}
Consider the differential equation
\begin{equation}\label{ded}
\phi'(z)=\frac{1}{\omega}[h(\phi(z))+gR(z)w(\phi(z))]
\end{equation}
where $h$,$w$ are defined by (\ref{spec}), $\omega>0$ is fixed,
and $R\in C^2[0,2\pi]$ is a positive function. Then there exists
{\bf{at most}} one value of $g$ for which the boundary-value
problem (\ref{ded}) and
\begin{equation}\label{inin}
\phi(0)=\pi,\;\;\phi(2\pi)=3\pi
\end{equation}
has a solution.
\end{lemma}

\noindent {\sc{proof:}} Assume by way of contradiction that
$g_2>g_1$ and  $\phi_1(z)$ and $\phi_2(z)$ satisfy the
differential equations
\begin{equation}\label{deg1}
\phi_1'(z)=\frac{1}{\omega}[ h(\phi_1(z))+ g_1 R(z)w(\phi_1(z))],
\end{equation}
\begin{equation}\label{deg2}
\phi_2'(z)=\frac{1}{\omega}[h(\phi_2(z))+ g_2 R(z)w(\phi_2(z))],
\end{equation}
with boundary conditions
\begin{equation}\label{ing12}
\phi_1(0)=\phi_2(0)=\pi.
\end{equation}
\begin{equation}\label{bwc}
\phi_1(2\pi)=\phi_2(2\pi)=3\pi.
\end{equation}
We define
$$\varphi(z)=\phi_2(z)-\phi_1(z).$$
From (\ref{ing12}) and (\ref{bwc}) we have
\begin{equation}\label{var}
\varphi(0)=\varphi(2\pi)=0.
\end{equation}
We now compute the first three derivatives of $\varphi(z)$ at
$z=0,2\pi$. We shall be using (\ref{hpi}), (\ref{wz}), as well as
the facts that
$$h'((2l+1)\pi)=0,\;\; w'((2l+1)\pi)=0\;\;\;\forall l\in \Integer,$$
$$h''((2l+1)\pi)=\beta-1,\;\;w''((2l+1)\pi)=1\;\;\;\forall l\in \Integer.$$
Substituting $z=0,2\pi$ into (\ref{deg1}),(\ref{deg2}) and using
(\ref{ing12}), (\ref{bwc}) we have
\begin{equation}\label{phid0}
\phi_1'(0)=\phi_2'(0)=
\phi_1'(2\pi)=\phi_2'(2\pi)=\frac{2}{\omega},
\end{equation}
hence
\begin{equation} \label{dvar}
\varphi'(0)=\varphi'(2\pi)=0.
\end{equation}
Differentiating (\ref{ded}) we have
\begin{equation}\label{degg}
\phi''(z)=\frac{1}{\omega} [h'(\phi(z))\phi'(z)+ g
R'(z)w(\phi(z))+ g R(z) w'(\phi(z))\phi'(z)],
\end{equation}
and substituting $z=0,2\pi$ into (\ref{degg}) and using
(\ref{ing12}),(\ref{bwc}) and (\ref{phid0}) we obtain
\begin{equation}\label{phidd0}
\phi_1''(0)=\phi_2''(0) =\phi_1''(2\pi)=\phi_2''(2\pi)=0,
\end{equation}
So that
\begin{equation}\label{ddvar}
\varphi''(0)=\varphi''(2\pi)=0.
\end{equation}
Differentiating (\ref{degg}) we have
\begin{eqnarray}\label{deggg}
\phi'''(z)&=&\frac{1}{\omega}[ h''(\phi(z))(\phi'(z))^2 +
h'(\phi(z))\phi''_{\omega,g}(z)\nonumber\\ &+& g
R''(z)w(\phi(z))+2 gR'(z) w'(\phi(z))\phi'(z)\nonumber\\&+&g
R(z)w''(\phi(z))(\phi'(z))^2 +
 g R(z)w'(\phi(z))\phi''(z)],
\end{eqnarray}
and upon substituting $z=0,2\pi$ into (\ref{deggg}) and using
(\ref{ing12}), (\ref{bwc}), (\ref{phid0}) and (\ref{phidd0}) we
obtain
$$\phi_1'''(0)=
\frac{4}{\omega^3}(\beta-1+g_1 R(0)),\;\;
\phi_2'''(0)=\frac{4}{\omega^3}(\beta-1+g_2 R(0))$$
$$\phi_1'''(2\pi)=
\frac{4}{\omega^3}(\beta-1+g_1 R(2\pi)),\;\;
\phi_2'''(2\pi)=\frac{4}{\omega^3}(\beta-1+g_2 R(2\pi))
$$ hence, using also the assumption that $g_2>g_1$,
\begin{equation}\label{dddvar}
\varphi'''(0)=\frac{4}{\omega^3}(g_2-g_1)R(0)>0.
\end{equation}
\begin{equation}\label{dddvar1}
\varphi'''(2\pi)=\frac{4}{\omega^3}(g_2-g_1)R(2\pi)>0.
\end{equation}
From (\ref{var}),(\ref{dvar}),(\ref{ddvar}), (\ref{dddvar}) and
(\ref{dddvar1}) we conclude that there exists $\epsilon>0$ such
that
\begin{equation}\label{po1}
z\in [0,\epsilon]\;\; \Rightarrow\;\; \varphi(z)>0
\end{equation}
\begin{equation}\label{po2}
z\in [2\pi-\epsilon,2\pi]\;\; \Rightarrow\;\; \varphi(z)<0.
\end{equation}
We will now use (\ref{po1}),(\ref{po2}) to derive a contradiction.
First note that if $z_0\in (0,2\pi)$ and $\varphi(z_0)=0$, then
using the definition of $\varphi(z)$ and lemma \ref{b1} we have
$$\pi<\phi_2(z_0)=\phi_1(z_0)<3\pi,$$
which implies
$$h(\phi_1(z_0))=h(\phi_2(z_0)),$$
$$w(\phi_1(z_0))=w(\phi_2(z_0))>0,$$
which together with (\ref{deg1}),(\ref{deg2}) and the assumption
that $g_2>g_1$ imply
$$\phi_2'(z_0)>\phi_1'(z_0),$$
or in other words that
$$\varphi'(z_0)>0.$$
We have thus shown that
\begin{equation}\label{k1} z\in
(0,2\pi),\;\;\varphi(z)=0\;\;\Rightarrow \varphi'(z)>0.
\end{equation}
Let
$$z_1=\inf \{ z\in (0,2\pi)\;|\; \varphi(z)<0\}$$
(note that by (\ref{po2}) the above set is nonempty, and by
(\ref{po1}) it is bounded from below. By (\ref{po1}) we have
$z_1>0$ and by continuity we have $\varphi(z_1)=0$. Hence by
(\ref{k1}) we have $\varphi'(z_1)>0$. But this implies that for
$z<z_1$ sufficiently close to $z_1$ we have $\varphi(z)<0$,
contradicting the definition of $z_1$. This contradiction
establishes the lemma.

\begin{question}
We have shown that several of the qualitative features that we
derived by direct computation in the case of uniform coupling
(section \ref{constant}) remain valid in the general case. It is
natural to ask whether more can be said, {\it{e.g.}}, whether for
any $K(x)$, when $\beta<0$ and $g>g_{crit}$ there exist
{\it{precisely}} two rotating waves. This would follow if we could
prove that the only local minimum of the function $g(\omega)$ is
the global one. However, as we show by numerical computations in
section \ref{smallwn}, there are in fact cases in which the
function $g(\omega)$ has {\it{two}} minima, and values of $g$ for
which {\it{four}} rotating waves exist. Thus a modified conjecture
is that for sufficiently large $g$ there are exactly two rotating
waves.
\end{question}

\begin{question}
Is it true that in the oscillatory case $\beta\geq 0$ the rotating
wave is always unique? In section \ref{constant} we saw that this
is the case when $K(x)$ is constant. This would follow if we could
show that $g(\omega)$ is an increasing function.
\end{question}

\section{Some quantitative bounds}
\label{quant}

Based on the results of the previous section, we now obtain some
explicit bounds for the velocities of the rotating waves and, in
the excitable case, for the critical synaptic-coupling strength
$g_{crit}$.

We first note that the inequalities (\ref{pp2}) imply that when
$c>0$
\begin{equation}\label{ask}
g_k(\omega)=\frac{1}{2\|K\|_{L^1[0,2\pi]}}(e^{\frac{c}{2}}-1)\omega^2+O(\omega)\;\;\;as\;\;\omega\rightarrow+\infty,
\end{equation}
while when $c=0$
\begin{equation}\label{ask0}
g_k(\omega)=\frac{\pi}{\|K\|_{L^1[0,2\pi]}}\omega+O(1)\;\;\;as\;\;\omega\rightarrow+\infty.
\end{equation}
These asymptotic expressions imply asymptotic expressions for the
velocity $\omega$ in the case $g\rightarrow \infty$ in the case
$\beta\geq 0$, and for the velocity of the fast wave in the case
$\beta<0$.

\begin{theorem}
\label{ssk}

\noindent (I) In the excitable case $\beta<0$, when $c>0$, the
velocity $\omega_f$ of the fast rotating wave satisfies
\begin{equation}\label{fwf}
\omega_f=\sqrt{\frac{2\|K\|_{L^1[0,2\pi]}}{e^{\frac{c}{2}}-1}}\sqrt{g}+O(1)\;\;\;as\;\;g\rightarrow
+\infty,
\end{equation}
while in the case $c=0$
\begin{equation}\label{fwf0}
\omega_f=\frac{\|K\|_{L^1[0,2\pi]}}{\pi}g+O(1)\;\;\;as\;\;g\rightarrow
+\infty,
\end{equation}

\noindent (II) In the case $\beta\geq 0$ we have the same
asymptotic formulas (\ref{fwf}), (\ref{fwf0}) for the velocity of
the rotating wave $\omega$.
\end{theorem}

By using the inequalities in (\ref{pp1}),(\ref{pp3}) and lemmas
\ref{solutions} and \ref{as} we can obtain explicit bounds on the
velocities of both the slow and fast rotating waves in terms of
the functions
$\omega_{c,\beta}$,$\underline{\omega}_{c,\beta}$,$\overline{\omega}_{c,\beta}$
defined by lemma \ref{solutions}, as well as simple asymptotic
bounds for the velocity of the slow wave as $g\rightarrow +\infty$
in the case $\beta<0$ (note that theorem \ref{ssk} above has
already provided us with asymptotics for the fast wave), and for
the velocity as $g\rightarrow -\infty$ in the case $\beta>0$.

\begin{theorem}
\label{ts}

\noindent (I) In the excitable case $\beta<0$, assume
$g>\frac{\Omega(c,\beta)}{\underline{K}}$. Then there exist a
`slow' rotating wave with velocity $\omega_s$ bounded from above
by
\begin{equation}\label{sw}
\underline{\omega}_{c,\beta}\Big(\overline{K}g\Big)\leq
\omega_s\leq \underline{\omega}_{c,\beta}\Big(\underline{K}g\Big)
\end{equation}
and a `fast' rotating wave with velocity $\omega_f$ bounded from
below by
\begin{equation}\label{fw}
\overline{\omega}_{c,\beta}\Big(\underline{K}g\Big) \leq \omega_f
\leq \overline{\omega}_{c,\beta}\Big(\overline{K}g\Big),
\end{equation}
where $\underline{\omega}_{c,\beta},\overline{\omega}_{c,\beta}$
are the functions defined by lemma \ref{solutions}.

\noindent As a consequence of (\ref{sw}) we have, for the slow
wave
\begin{equation}
\label{swe}
\frac{2|\beta|e^{\frac{c}{2}}}{\overline{K}}\frac{1}{g}+O\Big(\frac{1}{g^3}\Big)
\leq \omega_s\leq
\frac{2|\beta|e^{\frac{c}{2}}}{\underline{K}}\frac{1}{g}+O\Big(\frac{1}{g^3}\Big)
\;\;\;as\;\;g\rightarrow\infty.
\end{equation}

\noindent (II) In the oscillatory case $\beta>0$, there exists a
rotating wave solution for any value of $g\in \Real$, with
velocity $\omega$ bounded by
\begin{equation}\label{fw1}
\omega_{c,\beta}(\overline{K}g)\leq \omega \leq
\omega_{c,\beta}(\underline{K}g)\;\;\;for\;g\leq 0,
\end{equation}
\begin{equation}\label{fw2}
\omega_{c,\beta}(\underline{K}g)\leq \omega \leq
\omega_{c,\beta}(\overline{K}g)\;\;\;for\;g\geq 0,
\end{equation}
and we have the asymptotic inequalities
\begin{equation}\label{asv3301}
-\frac{4\beta
e^{\frac{c}{2}}}{\overline{K}}\frac{1}{g}+O\Big(\frac{1}{g^3}\Big)\leq
\omega\leq -\frac{4\beta
e^{\frac{c}{2}}}{\underline{K}}\frac{1}{g}+O\Big(\frac{1}{g^3}\Big)\;\;\;as\;\;g\rightarrow
-\infty.
\end{equation}

\noindent (III) In the borderline case $\beta=0$, there exists a
rotating wave solution for any value of $g>0$, with velocity
$\omega$ bounded as in (\ref{fw2}), and
$$\omega=o(1)\;\;\;as\;\; g\rightarrow 0.$$
\end{theorem}

We now derive explicit lower and upper bounds for the value of
$g_{crit}$ in the excitable case $\beta<0$, that is the critical
value of $g$ at which two rotating waves are born. Using
(\ref{gcrit}) and (\ref{pp1}) we obtain
$$g_{crit}=\min_{\omega>0}g(\omega)\geq \frac{1}{\overline{K}}
\min_{\omega>0}f_{c,\beta}(\omega) =
\frac{1}{\overline{K}}\Omega(c,\beta)
$$ and
$$g_{crit} =\min_{\omega>0}g(\omega) \leq \frac{1}{\underline{K}}\min_{\omega>0}
f_{c,\beta}(\omega)=\frac{1}{\underline{K}}\Omega(c,\beta),$$ so
that
\begin{lemma}\label{bgcrit} In the excitable case $\beta<0$
$$\frac{1}{\overline{K}}\Omega(c,\beta)\leq g_{crit} \leq
\frac{1}{\underline{K}}\Omega(c,\beta).$$
\end{lemma}

If, instead of using the inequality (\ref{pp1}) we use
(\ref{pp2}), we obtain
\begin{lemma}\label{bgcrit1} In the excitable case $\beta<0$
$$\frac{2}{\|K\|_{L^1[0,2\pi]}}\min_{\omega\geq 0}\Big(\frac{\omega^2}{4}-\beta
\Big)\Big(e^{\frac{c}{2}+\frac{2\pi}{\omega}}-1\Big)e^{-\frac{2\pi}{\omega}}
\leq g_{crit} \leq \frac{2}{\|K\|_{L^1[0,2\pi]}}\min_{\omega\geq
0}\Big(
\frac{\omega^2}{4}-\beta\Big)\Big(e^{\frac{c}{2}+\frac{2\pi}{\omega}}-
1\Big)$$
\end{lemma}

\section{Periodic travelling waves on the line}
\label{genper}

We now exploit the results about rotating waves on a ring obtained
in the previous sections  in order to derive results about
periodic waves on the line. As we have already noted in the
introduction, periodic waves with wave-number $k$ are the same as
rotating waves on the ring with $K=J_k$, where $J_k$ is defined by
(\ref{defjk}). We define the function
\begin{equation}\label{defromk}
R_{\omega,k}(z)=\int_{0}^{2\pi}{J_k(z-y)r_{\omega}(y)dy}
\end{equation}
as in (\ref{defrl}) (with $K=J_k$), and similarly the quantities
$$\overline{R}_{\omega,k}(z)=\sup_{x\in\Real}{R_{\omega,k}(x)},\;\;\;
\underline{R}_{\omega,k}=\inf_{x\in\Real}{R_{\omega,k}(x)}$$ as in
(\ref{defrlam}),
$$\overline{J}_k=\sup_{x\in\Real}{J_{k}(x)},\;\;
\underline{J}_k=\inf_{x\in\Real}{J_{k}(x)}$$ as in (\ref{defouk}),
and
$$\|J_k\|_{L^1[0,2\pi]}=\int_{0}^{2\pi}{J_k(x)dx}$$ as in (\ref{defk1}).

We note that by the definition (\ref{defjk}) of $J_k$ we have
\begin{equation}\label{l1}
\|J_k\|_{L^1[0,2\pi]}=\int_{-\infty}^{\infty}{J(z)dz}=\|J\|_{L^1(\Real)}.
\end{equation}

For each $\omega>0$, $k>0$, $g>0$ we define by
$\phi_{\omega,g,k}(z)$ the solution of the initial value problem
\begin{equation}\label{eqlk}
\phi_{\omega,g,k}'(z)=
\frac{1}{\omega}[h(\phi_{\omega,g,k}(z))+gR_{\omega,k}(z)w(\phi_{\omega,g,k}(z))],
\end{equation}
\begin{equation}\label{inlk}
\phi_{\omega,g,k}(0)=\pi,
\end{equation}
the function $\Psi_k(\omega,g)$ by
\begin{equation}\label{psik}
\Psi_k(\omega,g)=\frac{1}{3\pi}\phi_{\omega,g,k}(2\pi),
\end{equation}
and the frequency vs. synaptic-coupling strength curve for waves
of wave-number $k$ by
\begin{equation}\label{defsigk}
\Sigma_k=\{ (\omega,g)\;|\;\Psi_k(\omega,g)=1\}.
\end{equation}

In order to apply theorem \ref{curve} to $K=J_k$, we need a
condition ensuring that $J_k\in L^\infty(\Real)$. Such a condition
is given by the following

\begin{lemma}\label{kinf}
Assume $J$ satisfies (\ref{jdecay}). Then we have
$$|J_k(z)-\frac{1}{2\pi}\int_{-\infty}^{\infty}{J(x)dx}|\leq
\frac{1}{k}\|J'\|_{L^1(\Real)} \;\;\;\forall k>0,z\in\Real.$$ In
particular $J_k\in L^{\infty}(\Real)$.
\end{lemma}

\noindent {\sc{proof:}} We have
$$J_k'(x)=\frac{1}{k^2}\sum_{l=-\infty}^{\infty}{J'\Big(\frac{1}{k}(x-2\pi
l)\Big)},$$ hence
\begin{eqnarray}\label{inc}\|J_k'\|_{L^1[0,2\pi]}&=&\int_{0}^{2\pi}{|J_k'(x)|dx}\leq
\frac{1}{k^2}\sum_{l=-\infty}^{\infty}{\int_{0}^{2\pi}{|J'\Big(\frac{1}{k}(x-2\pi
l)\Big)|dx}}\nonumber\\&=&\frac{1}{k^2}\int_{-\infty}^{\infty}{|J'\Big(\frac{x}{k}\Big)|
dx}=\frac{1}{k}\int_{-\infty}^{\infty}{|J'(x)|
dx}=\frac{1}{k}\|J'\|_{L^1(\Real)}.
\end{eqnarray}
We have, for all $y,z\in [0,2\pi]$
$$|J_k(z)-J_k(y)|=|\int_y^z{J_k'(x)dx}|\leq
\int_{0}^{2\pi}{|J_k'(x)|dx}=\|J_k'\|_{L^1[0,2\pi]}.$$ By
periodicity of $J_k$ this inequality in fact holds for all
$y,z\in\Real$, and it implies
$$|J_k(z)-\frac{1}{2\pi}\int_0^{2\pi}{J_k(x)dx}|\leq \|J_k'\|_{L^1[0,2\pi]}
\;\;\;\forall x\in \Real.$$ Together with the equality (\ref{l1})
and (\ref{inc}), this gives the result of our lemma.

\vspace{0.4cm} From theorem \ref{curve} we thus obtain

\begin{theorem}\label{ptw}
Assume $J$ satisfies (\ref{jdecay}) and (\ref{jpo}). Then, for
each $k>0$, the curve $\Sigma_k$ defined by (\ref{defsigk}) can be
represented as the graph of a function:
\begin{equation}\label{repsigk}
\Sigma_k=\{(\omega,g_k(\omega)) \;|\; \omega\in (0,\infty)\},
\end{equation}
where $g_k:(0,\infty)\rightarrow \Real$ is a continuous function
which satisfies

\noindent (i) When $\omega^2\geq 4\beta$:
\begin{equation}\label{ppk1}\frac{1}{\overline{J}_k}f_{c,\beta}(\omega)\leq g_k(\omega)\leq
\frac{1}{\underline{J}_k}f_{c,\beta}(\omega),\end{equation}
\begin{equation}\label{ppk2}
\frac{2}{\|J\|_{L^1(\Real)}}\Big(\frac{\omega^2}{4}-\beta
\Big)\Big(e^{\frac{c}{2}+\frac{2\pi}{\omega}}-1\Big)e^{-\frac{2\pi}{\omega}}
\leq g_k(\omega) \leq \frac{2}{\|J\|_{L^1(\Real)}}\Big(
\frac{\omega^2}{4}-\beta\Big)\Big(e^{\frac{c}{2}+\frac{2\pi}{\omega}}-
1\Big).
\end{equation}

\noindent (ii) When $\omega^2\leq 4\beta$:
\begin{equation}\label{ppk3}\frac{1}{\underline{J}_k}f_{c,\beta}(\omega)\leq g_k(\omega)\leq
\frac{1}{\overline{J}_k}f_{c,\beta}(\omega),
\end{equation}
\begin{equation}\label{ppk4}
\frac{2}{\|J\|_{L^1(\Real)}}\Big(
\frac{\omega^2}{4}-\beta\Big)\Big(e^{\frac{c}{2}+\frac{2\pi}{\omega}}-
1\Big) \leq g_k(\omega) \leq \frac{2}{\|J\|_{L^1(\Real)}}\Big(
\frac{\omega^2}{4}-\beta\Big)\Big(e^{\frac{c}{2}+\frac{2\pi}{\omega}}-1\Big)e^{-\frac{2\pi}{\omega}}.
\end{equation}

\noindent and

\noindent (I) In the oscillatory case $\beta>0$
\begin{equation}\label{bb1}
\lim_{\omega\rightarrow 0+}g_k(\omega)=-\infty.
\end{equation}

\noindent (II) In the excitable case $\beta<0$
\begin{equation}\label{bb2}
\lim_{\omega\rightarrow 0+}g_k(\omega)=+\infty.
\end{equation}
\noindent (III) In the borderline case $\beta=0$
\begin{equation}\label{bb3}
\lim_{\omega\rightarrow 0+}g_k(\omega)=0.
\end{equation}
\end{theorem}

When $\beta<0$, for each $k>0$ we have the critical value
$$g_{crit}(k)=\min_{\omega>0}g_k(\omega),$$
above which we have the existence of two periodic travelling waves
of wave-number $k$. From lemma \ref{bgcrit1} we have

\begin{lemma}\label{bgcritk1}
Assume $J$ satisfies (\ref{jdecay}) and (\ref{jpo}), and
$\beta<0$. Then for all $k>0$
$$\frac{2}{\|J\|_{L^1(\Real)}}\min_{\omega\geq 0}\Big(\frac{\omega^2}{4}-\beta
\Big)\Big(e^{\frac{c}{2}+\frac{2\pi}{\omega}}-1\Big)e^{-\frac{2\pi}{\omega}}
\leq g_{crit}(k) \leq \frac{2}{\|J\|_{L^1(\Real)}}\min_{\omega\geq
0}\Big(\frac{\omega^2}{4}-\beta\Big)\Big(e^{\frac{c}{2}+\frac{2\pi}{\omega}}-
1\Big)$$
\end{lemma}
which shows that $g_{crit}(k)$ is bounded between two positive
constants for all $k>0$, as claimed in part (II)(B) of theorem
\ref{pertra}.

\vspace{0.4cm} Using theorem \ref{ssk}, the identity (\ref{l1}),
and the expression $v=\frac{\omega}{k}$ for the velocity of the
periodic travelling waves, we obtain the asymptotic velocity of
the periodic travelling wave in the oscillatory case, and of the
fast periodic travelling wave in the excitable case for the cse of
strong synaptic coupling ($g$).

\begin{theorem}
\label{ssk2} Assume $J$ satisfies (\ref{jdecay}) and (\ref{jpo}).
Then, for each $k>0$,

\noindent (I) In the excitable case $\beta<0$, when $c>0$, the
velocity $v_f$ of the fast periodic travelling wave of wave-number
$k$ satisfies
\begin{equation}\label{fwf1}
v_f=\frac{1}{k}\sqrt{\frac{2\|J\|_{L^1(\Real)}}{e^{\frac{c}{2}}-1}}\sqrt{g}+O(1)\;\;\;as\;\;g\rightarrow
+\infty,
\end{equation}
while in the case $c=0$
\begin{equation}\label{fwf01}
v_f=\frac{1}{k}\frac{\|J\|_{L^1(\Real)}}{\pi}g+O(1)\;\;\;as\;\;g\rightarrow
+\infty,
\end{equation}

\noindent (II) In the case $\beta\geq 0$ we have the same
asymptotic formulas (\ref{fwf}), (\ref{fwf0}) for the velocity of
the periodic travelling wave of wave-number $k$.
\end{theorem}

\vspace{0.4cm} Let us note that the quantitative bounds for the
velocities of travelling waves on a ring given in theorem
\ref{ts}, imply bounds for the velocities of periodic travelling
waves of wave-number $k$, and asymptotic bounds as $g\rightarrow
\infty$ for the velocity of obtained by replacing
$\overline{K},\underline{K}$ by $\overline{J}_k, \underline{J}_k$.
However in contrast with the result of theorem \ref{ssk2}, these
bounds depend on the quantities $\overline{J}_k, \underline{J}_k$,
so the dependence on $k$ is not as explicit and must be computed
for each individual synaptic-coupling kernel $J$.

\vspace{0.4cm} The curves $\Sigma_k$ obviously depend on the shape
of the synaptic-coupling kernel $J$. However in the next sections,
where we investigate the limits of large and of small wave number,
we shall discover that in these limits these curves tend to shapes
which are independent of the shape of $J$ (depending on $J$ only
through the norm $\|J\|_{L^1(\Real)})$.

\section{Periodic travelling waves on the line: large wave-number}
\label{large}

This section is devoted to the study of periodic travelling waves
with large wave-number $k$. The main point which we shall prove is
that although clearly the frequency vs. coupling-strength curve
for each finite $k>0$ depends on the details of the
coupling-kernel $J$, as $k\rightarrow \infty$ this curve
approaches a limiting curve which is independent of details of
$J$, depending only (in a trivial way) on the norm
$\|J\|_{L^1(\Real)}$, and is thus `universal'. Indeed we shall see
that the case $k\rightarrow \infty$ is related to the case of
uniform coupling on a ring studied in section \ref{constant}, and
thus we will be able to obtain a closed expression for this
limiting curve.

\vspace{0.4cm}

From lemma \ref{kinf} we have
$$\frac{1}{2\pi}\|J\|_{L^1(\Real)}-\frac{1}{k}\|J'\|_{L^1(\Real)}\leq
\underline{J}_k\leq \overline{J}_k\leq
\frac{1}{2\pi}\|J\|_{L^1(\Real)}+\frac{1}{k}\|J'\|_{L^1(\Real)},$$
implying
$$\lim_{k\rightarrow+\infty}{\underline{J}_k}=
\lim_{k\rightarrow+\infty}{\overline{J}_k}=\frac{1}{2\pi}\|J\|_{L^1(\Real)},$$
which, together with (\ref{ppk1}) and (\ref{ppk3}) implies

\begin{theorem}\label{lkl0} Assume $J$ satisfies (\ref{jdecay}) and (\ref{jpo}). We have
$$\lim_{k\rightarrow +\infty}g_k(\omega)=g_{\infty}(\omega),$$
where
$$g_{\infty}(\omega)=\frac{2\pi}
{\|J\|_{L^1(\Real)}}f_{c,\beta}(\omega)$$ uniformly for $\omega$
in compact subsets of $(0,\infty)$, and if $\beta<0$
$$\lim_{k\rightarrow +\infty}g_{crit}(k)=
\frac{2\pi} {\|J\|_{L^1(\Real)}}\Omega(c,\beta).$$
\end{theorem}

This implies

\begin{theorem}
\label{lk11} Assume $J$ satisfies (\ref{jdecay}) and (\ref{jpo}).
Then

\noindent (I) In the oscillatory case $\beta>0$, for any
$g\in\Real$ and $k>0$ we have: there exists a periodic travelling
wave with wave-number $k$, and its frequency satisfies
\begin{equation}\label{bop}
\lim_{k\rightarrow\infty}{\omega(k)}=\omega_{c,\beta}\Big(\frac{g}{2\pi}\|J\|_{L^1(\Real)}\Big),
\end{equation}
where $\omega_{c,\beta}$ is given by lemma \ref{solutions}.

\noindent (II) In the excitable case $\beta<0$:

\noindent (i) If $g<\frac{2\pi}
{\|J\|_{L^1(\Real)}}\Omega(c,\beta)$ then for $k$ sufficiently
large there are no periodic travelling waves with wave-number $k$.

\noindent (ii) If $g>\frac{2\pi}
{\|J\|_{L^1(\Real)}}\Omega(c,\beta)$ then for $k$ sufficiently
large there are two periodic travelling waves with wave-number
$k$, and their frequencies $\omega_s(k)$, $\omega_f(k)$ satisfy
$$\lim_{k\rightarrow\infty}{\omega_s(k)}=
\underline{\omega}_{c,\beta}\Big(\frac{g}{2\pi}\|J\|_{L^1(\Real)}\Big),
\;\;\;\lim_{k\rightarrow\infty}{\omega_f(k)}
=\overline{\omega}_{c,\beta}\Big(\frac{g}{2\pi}\|J\|_{L^1(\Real)}\Big),$$
where the functions
$\underline{\omega}_{c,\beta},\overline{\omega}_{c,\beta}$ are
given by lemma \ref{solutions}.

\noindent (III) In the boundary case $\beta=0$:

\noindent (i) For any $g>0$ we have: for any $k>0$ there exists a
periodic travelling wave with wave-number $k$, and its frequency
satisfies (\ref{bop}).

\noindent (ii) For any $g\leq 0$ there are no periodic travelling
waves.
\end{theorem}

\begin{figure}
\centering
    \includegraphics[height=7cm,width=7cm, angle=0]{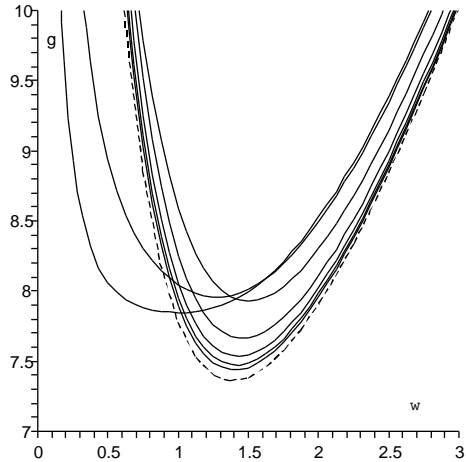}
    \caption{Graphs of the functions $g_k(\omega)$ for $k=0.25,0.5,1.5,2.5,3.5,4.5,5.5$, and the graph of the function $g_{\infty}(\omega)$ (dashed), in the case $\beta=-0.5$,$c=1$ obtained numerically.}
    \label{grgk}
\end{figure}

\vspace{0.4cm} We now present the results of some numerical
computations which we carried out, and these will demonstrate the
results of theorem \ref{lkl0}. For our computations we chose
$$J(x)=e^{-|x|},$$
so that $\|J\|_{L^1(\Real)}=2$ and a calculation shows that
$$J_k(z)=\frac{1}{k}
\Big[\frac{e^{-\frac{z}{k}}}{1-e^{-\frac{2\pi}{k}}}+\frac{e^{\frac{z}{k}}}{e^{\frac{2\pi}{k}}-1}\Big]
,\;\;\;0<z<2\pi.$$ We use this expression to compute
$R_{\omega,k}(z)$ according to \ref{defromk}, obtaining
$$R_{\omega,k}(z)=\frac{\omega^2}{k^2-\omega^2}\{\frac{1-e^{-\frac{2\pi}{\omega}}}
{e^{\frac{2\pi}{k}}-1}\Big[\frac{k}{\omega}\Big(e^{\frac{2\pi-z}{k}}+e^{\frac{z}{k}}\Big)+
\Big(e^{\frac{2\pi-z}{k}}-e^{\frac{z}{k}}\Big)\Big]-\frac{2}{k}e^{-\frac{z}{\omega}}\},$$
and then solve the initial value problem (\ref{eqlk}),(\ref{inlk})
numerically to obtain the function $\Psi_k(\omega,g)$ defined by
(\ref{psik}), and finally we solve the implicit equation
$\Psi_k(\omega,g_k(\omega))=1$ numerically to obtain the function
$g_k(\omega)$. The resulting graphs, for
$k=0.25,0.5,1.5,2.5,3.5,4.5,5.5$ are shown in figure \ref{grgk},
together with the graph of $g_{\infty}(\omega)$, plotted as a
dashed line. One can see that $g_k(\omega)$ approaches
$g_{\infty}(\omega)$ as $k$ increases, as claimed by theorem
\ref{lkl0}.

\section{Periodic travelling waves on the line: small wave-number}
\label{smallwn}

In this section we study the limit $k\rightarrow 0$. As in the
case of $k\rightarrow \infty$ studied in the previous section, we
shall see that the frequency vs. coupling-strength curve
approaches a `universal' limit independent of the details of the
synaptic kernel. Unlike in the previous section, we shall not be
able to derive a closed expression for this limiting curve, but we
shall be able to derive some of its key properties, which lead to
some interesting results about the small wave-number limit.

We will need a further assumption on the decay of $J$, namely
\begin{equation}\label{decayj}
zJ(z)\in L^1(\Real).
\end{equation}

\begin{lemma}\label{delta}
Assume $J(x)\in L^1(\Real)$ and (\ref{decayj}) holds. Assume
$f:\Real\rightarrow \Real$ is $2\pi$-periodic and continuous and
f(0)=0. Then
$$\lim_{k\rightarrow 0}{\frac{1}{k}\int_{-\infty}^{\infty}{J\Big(
\frac{x}{k}\Big)f(x)dx}}=0.$$
\end{lemma}

\noindent {\sc{proof:}} By a density argument it is sufficient to
prove the claim assuming that $f$ is $C^1$. This assumption and
the periodicity of $f$ implies that there exists a constant $C>0$
such that
$$|f(x)|\leq C|x|\;\;\;\forall x\in\Real.$$
Thus
\begin{eqnarray*}
|\frac{1}{k}\int_{-\infty}^{\infty}{J\Big(\frac{x}{k}\Big)f(x)dx}|
\leq C
\frac{1}{k}\int_{-\infty}^{\infty}{J\Big(\frac{x}{k}\Big)|x|dx}= C
k \int_{-\infty}^{\infty}{J(v)|v|dv}
\end{eqnarray*}
which implies claim of the lemma.

\begin{lemma}\label{skrl}
Assume $J$ satisfies (\ref{jdecay}),(\ref{jpo}) and
(\ref{decayj}). Then
\begin{equation}\label{limi}
\lim_{k\rightarrow
0}{R_{\omega,k}(z)}=r_{\omega}(z)\|J\|_{L^1(\Real)},
\end{equation}
uniformly in $z\in \Real$ and $\omega$ in compact subsets of
$(0,\infty)$.
\end{lemma}

\noindent {\sc{proof:}} (i) We note that, using the
$2\pi$-periodicity of $J_k$ and $r_{\omega}$,
\begin{eqnarray}\label{iden1}&&\int_{0}^{2\pi}{J_k(z-x)r_{\omega}(x)dx}=
\int_{0}^{2\pi}{J_k(x)r_{\omega}(z-x)dx}\\
&=&\frac{1}{k}\sum_{l=-\infty}^{\infty}{\int_{0}^{2\pi}
J\Big(\frac{1}{k}(x-2\pi l)\Big)r_{\omega}(z-x)dx}\nonumber\\
&=&\frac{1}{k}\sum_{l=-\infty}^{\infty}{\int_{2l\pi}^{2(l+1)\pi}
J\Big(\frac{x}{k}\Big)r_{\omega}(z-x)dx}=\frac{1}{k}\int_{-\infty}^{\infty}{
J\Big(\frac{x}{k}\Big)r_{\omega}(z-x)dx}\nonumber
\end{eqnarray}
We therefore have
\begin{eqnarray*}&\|& R_{\omega,k}(z) -
r_{\omega}(z)\|J\|_{L^1(\Real)}\|_{L^1[0,2\pi]}\\&=&
\|\int_{0}^{2\pi}{J_k(z-x)r_{\omega}(x)dx}-r_{\omega}(z)\int_{-\infty}^{\infty}{J(x)dx}\|_{L^1[0,2\pi]}
\\&\leq&
\frac{1}{k}\int_{0}^{2\pi}\int_{-\infty}^{\infty}J\Big(\frac{x}{k}\Big)
|r_{\omega}(z-x)-r_{\omega}(z)|dx dz \\&=&
\frac{1}{k}\int_{-\infty}^{\infty}J\Big(\frac{x}{k}\Big)
\int_{0}^{2\pi}|r_{\omega}(z-x)-r_{\omega}(z)|dz dx=
\frac{1}{k}\int_{-\infty}^{\infty}J\Big(\frac{x}{k}\Big) f(x) dx
\end{eqnarray*}
where
$$f(x)=\int_{0}^{2\pi}|r_{\omega}(z-x)-r_{\omega}(z)|dz.$$
Noting that $f$ is continuous and $f(0)=0$, this proves that
(\ref{limi}) holds in the sense of convergence in $L^1[0,2\pi]$,
uniformly for $\omega$ in compact subsets of $(0,\infty)$.

\noindent (ii) Applying the identity (\ref{iden1}) with $J$
replaced by $J'$ we obtain
$$R_{\omega,k}'(z)=\int_0^{2\pi}{J'_k(z-x)r_{\omega}(x)dx}=\frac{1}{k}
\int_{-\infty}^{\infty}{J'\Big(\frac{x}{k}\Big)r_{\omega}(z-x)dx}$$
so that
$$|R_{\omega,k}'(z)|\leq \|r_{\omega}\|_{L^{\infty}(\Real)}\|J'\|_{L^1(\Real)},\;\;\;\forall
z\in\Real,$$ so that the family $\{R_{\omega,k}\;|\; \omega\in
I,\;k>0\}$, where $I\subset (0,\infty)$ is compact, is uniformly
Lipschitz, and a standard argument using the Arzeli-Ascola theorem
and part (i) of the proof implies that (\ref{limi}) holds in the
sense of uniform convergence. \vspace{0.4cm}

This uniform convergence, and standard results on continuity of
differential equations with respect to parameters implies that
$$\lim_{k\rightarrow
0}{\phi_{\omega,g,k}(z)}=\phi_{\omega,g,0}(z),$$ uniformly for
$z\in [0,2\pi]$ and $(\omega,g)$ in compact subsets of
$(0,\infty)\times \Real$, where $\phi_{\omega,g,k}$ is defined by
(\ref{eqlk}),(\ref{inlk}) for $k>0$, and $\phi_{\omega,g,0}$ is
defined by
\begin{equation}\label{pl0} \phi_{\omega,g,0}'(z)=\frac{1}{\omega}[
h(\phi_{\omega,g,0}(z)) + g \|J\|_{L^1(\Real)}
r_{\omega}(z)w(\phi_{\omega,g,0}(z))],
\end{equation}
\begin{equation}\label{pl1}
\phi_{\omega,g,0}(0)=\pi.
\end{equation}
This in turn implies that
\begin{lemma}\label{psik0}
Assume $J$ satisfies (\ref{jdecay}), (\ref{jpo}) and
(\ref{decayj}). Then
$$\lim_{k\rightarrow 0}\Psi_{k}(\omega,g)=\Psi_0(\omega,g)$$
uniformly for $(\omega,g)$ in compact subsets of
$(0,\infty)\times\Real$, where $\Psi_k$ is defined by (\ref{psik})
and $\Psi_0$ is defined by
\begin{equation}\label{defpsi0}
\Psi_0(\omega,g)=\frac{1}{3\pi}\phi_{\omega,g,0}(2\pi).
\end{equation}
\end{lemma}

We study the set
\begin{equation}\label{defsig0}
\Sigma_0=\{ (\omega,g)\;|\; \Psi_0(\omega,g)=1\},
\end{equation}
that is, the limiting curve of the frequency vs. coupling strength
curves $\Sigma_k$ ($k>0$).

\begin{lemma}\label{eg0}
Assume $J$ satisfies (\ref{jdecay}), (\ref{jpo}) and
(\ref{decayj}). There exists a continuous function
$g_0:(0,\infty)\rightarrow \Real$ such that
$$\Sigma_0=\{(\omega,g_0(\omega))\;|\; \omega>0\},$$
and we have
\begin{equation}\label{gkk}
\lim_{k\rightarrow 0}g_k(\omega)=g_0(\omega),
\end{equation}
uniformly for $\omega$ in compact subsets of $(0,\infty)$.
\end{lemma}

\noindent {\sc{proof:}} Using lemma \ref{unig}, with
$R(z)=\|J\|_{L^1(\Real)} r_{\omega}(z)$, we conclude that for each
$\omega>0$ there exists {\bf{at most one}} value of $g$ for which
$(\omega,g)\in \Sigma_0$. To show that such a value of $g$ indeed
exists for any $\omega>0$ we need only note that, fixing
$\omega>0$, the set of numbers $\{g_k(\omega)\;|\; k>0\}$ is
bounded by (\ref{ppk2}) and (\ref{ppk4}), hence we can find a
subsequence $\{ k_i \}_{i=1}^{\infty}$, with $k_i\rightarrow 0$ as
$i\rightarrow \infty$ so that
$g^*=\lim_{i\rightarrow\infty}{g_{k_i}(\omega)}$ exists. By lemma
\ref{psik0} this implies $\Psi_0(g^*,\omega)=1$, as we wished to
show.

To prove (\ref{gkk}) we fix a closed interval $I\subset
(0,\infty)$. Assume by way of contradiction that (\ref{gkk}) does
not hold uniformly in $I$. This means that there exists an
$\epsilon>0$ and sequences $\{k_i\}_{i=1}^{\infty}\subset
(0,\infty)$ with $\lim_{i\rightarrow\infty}k_i=0$ and
$\{\omega_i\}_{i=1}^{\infty}\subset I$ with
\begin{equation}\label{contra}
|g_{k_i}(\omega_i)-g_0(\omega_i)|>\epsilon\;\;\;\forall i.
\end{equation}
By taking a subsequence we may assume that the sequence $\omega_i$
converges, say to $\omega$. By (\ref{ppk2}) and (\ref{ppk4}) the
sequence $g_{k_i}(\omega_i)$ is bounded, so by taking a
subsequence again we may assume that $g_{k_i}(\omega_i)\rightarrow
g^*$ as $i\rightarrow\infty$. From (\ref{contra}) we obtain
\begin{equation}\label{contra1}
|g^*-g_0(\omega)|>\epsilon
\end{equation}
for all $i$ sufficiently large. On the other hand we have
$\Psi_{k_i}(g_{k_i}(\omega),\omega_i)=1$ for all $i$, which going
to the limit $i\rightarrow \infty$ implies $\Psi_0(g^*,\omega)=1$,
so that $g^*=g_0(\omega)$, contradicting (\ref{contra1}), and
completing the proof.

\vspace{0.4cm}
Our aim now is to study the function $g_0(\omega)$,
whose properties will enable us to deduce results about the small
wave-number limit.

\begin{lemma}\label{pg0}
Assume $J$ satisfies (\ref{jdecay}), (\ref{jpo}) and
(\ref{decayj}). We have
\begin{equation}\label{g0inf}
\lim_{\omega\rightarrow +\infty}g_0(\omega)=+\infty
\end{equation}
and

\noindent (I) If $\beta>0$ then
\begin{equation}\label{cc1}
\lim_{\omega\rightarrow 0+}g_0(\omega)=-\infty.
\end{equation}

\noindent (II) If $\beta<0$ then
\begin{equation}\label{cc2}
g_0(0)=\lim_{\omega\rightarrow 0+}g_0(\omega)
\end{equation}
exists
and is a finite positive number.
\end{lemma}

We note that (\ref{g0inf}) follows at once from (\ref{gkk}) and
the inequality (\ref{ppk2}). The rest of lemma \ref{pg0} will be
proven below.

Let us note the fact that in the case $\beta<0$ the above lemma
shows that $g_0(\omega)$ is qualitatively different from
$g_k(\omega)$ ($k>0$): compare (\ref{bb2}) and (\ref{cc2}). This
has some interesting consequences for the velocity of the slow
wave as $k\rightarrow 0$, as we see in the following theorem,
which follows at once from lemmas \ref{eg0} and \ref{pg0}.

\begin{theorem}\label{smallkt0}
Assume $J$ satisfies (\ref{jdecay}), (\ref{jpo}) and
(\ref{decayj}).

\noindent (I) If $\beta>0$ then for all $g$ and any $k>0$ there
exists at least one periodic travelling wave with wave-number $k$,
and as $k\rightarrow 0$ the frequencies of these waves approach
some $\omega>0$, which satisfies
\begin{equation}\label{psi02}
g_0(\omega)=g.
\end{equation}

\noindent (II) Assume $\beta<0$. Let
$$g_{crit}(0)=\inf_{\omega>0}{g_0(\omega)}.$$
Then $$\lim_{k\rightarrow 0} g_{crit}(k)=g_{crit}(0)$$ and

\noindent (i) If $g<g_{crit}(0)$ then for sufficiently small $k>0$
there are no periodic travelling waves with wave-number $k$.

\noindent (ii) If $g_{crit}(0)<g<g_0(0)$ then for sufficiently
small $k>0$ there are at least two periodic travelling waves with
wave-number $k$,  their frequencies approaching solutions of
(\ref{psi02}) as $k\rightarrow 0$.

\noindent (iii) If $g>g_0(0)$ then for sufficiently small $k>0$
there are at least two periodic travelling waves with wave-number
$k$. As $k\rightarrow 0$, the frequency of the fast wave
approaches a solution of (\ref{psi02}), while the frequency of the
slow wave approaches $0$.
\end{theorem}

We now turn to the proof of lemma \ref{pg0}.

Let us consider the differential equation
\begin{equation}\label{dlim}
\varphi_A'(z)=h(\varphi_A(z))+Ae^{-z}w(\varphi_A(z))
\end{equation}
with initial condition
\begin{equation}\label{inlim}
\varphi_A(0)=\pi.
\end{equation}

\begin{lemma}\label{aid1}
For any $A\in\Real$ there is {\bf{at most one}} value of $z$ for
which $\varphi_A(z)=3\pi$.
\end{lemma}

\noindent {\sc{proof:}} From (\ref{dlim}), (\ref{hpi}) and
(\ref{wz}) we have
$$\varphi_A(z)=3\pi\;\;\Rightarrow\;\; \varphi_A'(z)=2.$$
Thus, lemma \ref{cut} implies the result.

\vspace{0.4cm} We note that if we define the function
$A:(0,\infty)\rightarrow\Real$ by
\begin{equation}\label{aspe}
A(\omega)=\|J\|_{L^1(\Real)}\Upsilon(\omega)g_0(\omega),\end{equation}
where $\Upsilon(\omega)$ is defined by (\ref{defups}), then
\begin{equation}\label{re1}
\phi_{\omega,g_0(\omega),0}(z)=\varphi_{A(\omega)}\Big(\frac{z}{\omega}
\Big),
\end{equation}
and therefore
\begin{equation}\label{id0}
\varphi_{A(\omega)}\Big(\frac{2\pi}{\omega}\Big)=
\phi_{\omega,g_0(\omega),0}(2\pi)=3\pi\;\;\;\forall \omega>0.
\end{equation}
We note that (\ref{id0}) can be regarded as an alternative
(implicit) definition of the function $A(\omega)$.

Let us note a few properties of the function $\Upsilon$ which will
be useful, and which follow from its definition by elementary
calculus
\begin{lemma}\label{prups}
$\Upsilon$ is an increasing concave function with
\begin{equation}\label{ups0}
\Upsilon(0)>0,
\end{equation}
\begin{equation}\label{ups1}
\Upsilon'(0)>0
\end{equation} and
\begin{equation}\label{ups2}
\lim_{\omega\rightarrow +\infty}\Upsilon(\omega)<\infty.
\end{equation}
\end{lemma}

The next lemma states some key properties of the function
$A(\omega)$.

\begin{lemma}\label{conj}
$A(\omega)$ is an increasing function, and

\noindent (I) If $\beta>0$ then
\begin{equation}\label{a0}
\lim_{\omega\rightarrow 0+}A(\omega)=-\infty,
\end{equation}

\noindent (II) If $\beta<0$ then
\begin{equation}\label{a01}
\lim_{\omega\rightarrow 0+}A(\omega)=A^*>0,
\end{equation}
\end{lemma}

\noindent {\sc{proof:}} From (\ref{g0inf}) and (\ref{aspe}) we
have
\begin{equation}\label{Ainf}
\lim_{\omega\rightarrow +\infty}A(\omega)=+\infty,
\end{equation}
so that to prove that $A(\omega)$ is increasing it suffices to
prove that it is one-to-one. Assume then that
$A(\omega_1)=A(\omega_2)=A_0$. By (\ref{id0}) we have
$$\varphi_{A_0}\Big( \frac{2\pi}{\omega_1}\Big)=
\varphi_{A_0}\Big( \frac{2\pi}{\omega_2}\Big)=3\pi.$$ But by lemma
\ref{aid1} this implies $\omega_1=\omega_2$, so we have proven
that $A(\omega)$ is one-to-one.

If we assume that $\beta>0$, then it is easy to see that for any
$A\in \Real$ there exists $z>0$ for which $\varphi_{A}(z)=3\pi$.
In other words, $A((0,\infty))=\Real$. Combining this with the
fact that $A(\omega)$ is continuous and increasing and with
(\ref{Ainf}) implies (\ref{a0}).

We now assume $\beta<0$. In this case it is easy to see that there
exists some $A_1>0$ such that
$$A<A_1\;\;\Rightarrow\;\; \varphi_{A}(z)<2\pi\;\;\forall z>0.$$
Therefore (\ref{id0}) implies that $A(\omega)\neq A_1$ for all
$\omega>0$, and by (\ref{Ainf}) this implies $A(\omega)>A_1$ for
all $\omega>0$. Thus $A(\omega)$ is bounded from below, and since
we have already proven that it is an increasing function we
conclude that the limit $A^*$ in (\ref{a01}) exists and $A^*\geq
A_1>0$.

\vspace{0.4cm} Lemma \ref{pg0} follows at once from (\ref{aspe})
and lemmas \ref{prups} and \ref{conj}.

\begin{figure}
\centering
    \includegraphics[height=7cm,width=7cm, angle=0]{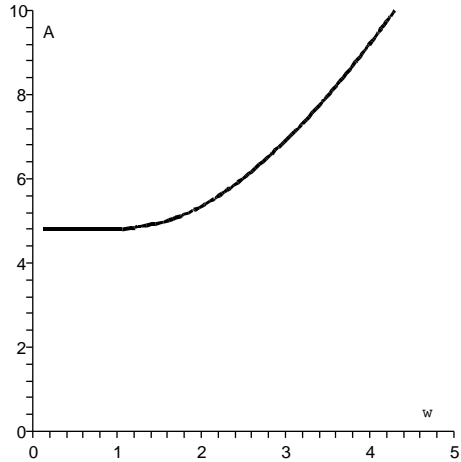}
    \caption{Graph of the function $A(\omega)$ defined by (\ref{id0}), in the case $\beta=-0.5$, obtained numerically.}
    \label{Aomega}
\end{figure}

\vspace{0.4cm} We now note two facts which are apparent when
plotting the graph of $A(\omega)$ (see figure \ref{Aomega} for the
graph of $A(\omega)$ when $\beta=-0.5$), obtained numerically, but
which we have not been able to prove analytically, so that they
retain the status of conjectures. Assuming these conjectures to be
valid, we shall be able to obtain some more information about the
function $g_0(\omega)$.

\begin{conjecture}\label{co1}
If $\beta<0$ then
\begin{equation}\label{ad0}
\lim_{\omega\rightarrow 0}A'(\omega)=0.
\end{equation}
\end{conjecture}

\begin{conjecture}\label{co2}
If $\beta<0$ then $A(\omega)$ is a convex function.
\end{conjecture}

\begin{lemma}\label{geni}
Let
\begin{equation}\label{psi03}
g(\omega)=\frac{A(\omega)}{\Upsilon(\omega)},\;\;\;\omega>0
\end{equation}
where A is continuous function satisfying (\ref{a0}), (\ref{a01}),
(\ref{ad0}), and $\Upsilon$ is continuous, increasing and concave
and satisfies (\ref{ups0}),(\ref{ups1}),(\ref{ups2}). Then the
function $g$ has a global minimum at $\omega_0>0$, and satisfies
\begin{equation}\label{limga}
\lim_{\omega\rightarrow\infty}{g(\omega)}=+\infty.
\end{equation}
If, moreover, $A$ is convex, then $g$ is decreasing on
$[0,\omega_0)$ and increasing on $(\omega_0,\infty)$.
\end{lemma}

\noindent {\sc{proof:}} We note first that (\ref{limga}) follows
from (\ref{a0}),(\ref{ups2}). We have
$$g'(\omega)=\frac{A'(\omega)\Upsilon(\omega)-A(\omega)\Upsilon'(\omega)}
{(\Upsilon(\omega))^2},$$ so (\ref{a01}),(\ref{ad0}) and
(\ref{ups1}) imply that
\begin{equation}\label{dls}
g'(0)<0,
\end{equation}
so $g$ is decreasing near $\omega=0$. Therefore the global minimum
of $g$ is attained at some $\omega_0>0$.

Assuming now that $A$ is convex, we have
$$g''(\omega)=\frac{A''(\omega)\Upsilon(\omega)-A(\omega)\Upsilon''(\omega)
}{(\Upsilon(\omega))^2}+\frac{2\Upsilon'(\omega)
[A'(\omega)\Upsilon(\omega)-A(\omega)\Upsilon'(\omega)]}{(\Upsilon(\omega))^3}.$$
The first term above is always positive because $A$ is convex and
$\Upsilon$ is concave. This implies the following key property of
$g$:
\begin{equation}
\label{kpop} g'(\omega)\geq 0\;\;\Rightarrow\;\;g''(\omega)>0.
\end{equation}
We note that by (\ref{limga}) there must exist some $\omega>0$
with $g'(\omega)>0$. Let us define
$$\omega_0=\inf \{ \omega>0 \;|\; g'(\omega)\geq 0 \}.$$
By (\ref{dls}), we have $\omega_0>0$. From (\ref{kpop}) it easily
follows that that if $\omega>\omega_0$ then $g'(\omega)>0$. By
definition of $\omega_0$ we have $g'(\omega)<0$ for all $\omega\in
[0,\omega_0)$. The lemma is thus established.

\vspace{0.4cm}
We thus have the following results

\begin{lemma}\label{conj1}
Assume $J$ satisfies (\ref{jdecay}), (\ref{jpo}) and
(\ref{decayj}). Assume that conjecture \ref{co1} above is valid.
Then when $\beta<0$, $g_0(\omega)$ has a global minimum at some
$\omega_0>0$.
\end{lemma}

\begin{lemma}\label{conj2}
Assume $J$ satisfies (\ref{jdecay}), (\ref{jpo}) and
(\ref{decayj}). Assume that both conjectures  \ref{co1} and
\ref{co2} above are valid. Then when $\beta<0$, $g_0(\omega)$ has
a global minimum at some $\omega_0>0$, and is decreasing on
$(0,\omega_0)$ and increasing on $(\omega_0,\infty)$.
\end{lemma}

\begin{figure}
\centering
    \includegraphics[height=7cm,width=7cm, angle=0]{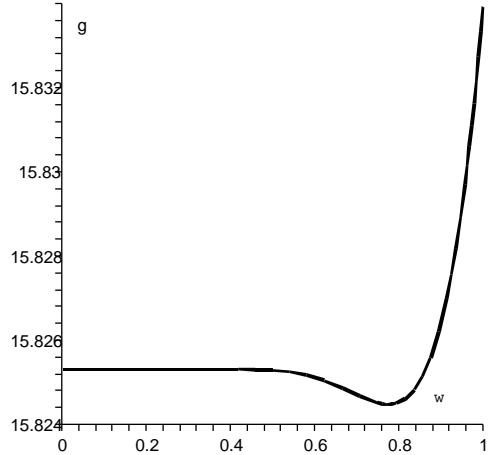}
    \caption{Graph of the function $g_0(\omega)$ in the case $J(z)=e^{-|z|}$, $\beta=-0.5$, $c=1$, obtained numerically.}
    \label{gig}
\end{figure}

In figure \ref{gig} we plot the function $g_0(\omega)$, when
$\beta=-0.5$ and $c=1$, obtained by solving the equation
$\Psi_0(\omega,g_0(\omega))=1$ for $g_0(\omega)$ numerically,
where $\Psi_0$ is defined by (\ref{defpsi0}). The qualitative
properties claimed in lemma \ref{conj2} can be seen in this plot.

\vspace{0.4cm}
Thus, in the case $\beta<0$, when $g$ is
sufficiently large the frequencies of the fast and the slow
travelling waves approach two positive values as $k\rightarrow 0$
(part (II)(ii) of the lemma), while if
\begin{equation}\label{sm1}
g_{crit}(0)<g_0(0)
\end{equation}
part (II)(iii) of theorem \ref{smallkt0} implies that we have a
range of values of $g$ for which the frequency of the slow wave
approaches $0$ as $k\rightarrow 0$. We note that, by definition,
$g_{crit(0)}\leq g_0(0)$, so to show that case (ii) corresponds to
a non-empty interval of values of $g$ we need only to show that
the last inequality is strict. This follows at once from the
conclusion of lemma \ref{conj1}, so that it is true modulo the
validity of conjecture \ref{co1}.

Furthermore if we assume that both conjectures \ref{co1} and
\ref{co2} are valid, then lemma \ref{conj2} tells us that
$g_0(\omega)$ is unimodal, so we obtain the following
strengthening of theorem \ref{smallkt0}:

\begin{theorem}\label{smallkt}
Assume $J$ satisfies (\ref{jdecay}), (\ref{jpo}) and
(\ref{decayj}). Suppose that conjectures \ref{co1} and \ref{co2}
are valid. Assume $\beta<0$, and let
$g_{crit}(0)=\min_{\omega>0}{g_0(\omega)}$. Then
$g_0(0)>g_{crit}(0)$, and

\noindent (i) If $g<g_{crit}(0)$ then for sufficiently small $k>0$
there are no periodic travelling waves with wave-number $k$.

\noindent (ii) If $g_{crit}(0)<g<g_0(0)$, then (\ref{psi02}) has
precisely two solutions
$0<\underline{\omega}<\overline{\omega}<\infty$, and for
sufficiently small $k>0$ there are at least two periodic
travelling waves with wave-number $k$ with the frequencies of the
slow and fast waves  approaching $\underline{\omega}$ and
$\overline{\omega}$ as $k\rightarrow 0$.

\noindent (iii) If $g>g_0(0)$ then (\ref{psi02}) has a unique
solution $\omega$, and for sufficiently small $k>0$ there are at
least two periodic travelling waves with wave-number $k$, and the
frequency of the slow and fast waves approaches $0$ and $\omega$,
respectively, as $k\rightarrow 0$.
\end{theorem}

\begin{figure}
\centering
    \includegraphics[height=7cm,width=7cm, angle=0]{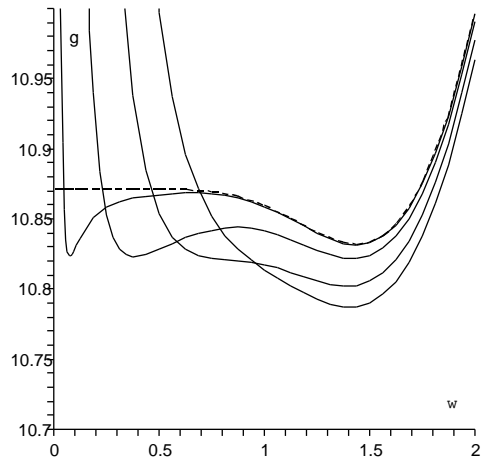}
    \caption{Numerically computed graphs of the functions $g_k(\omega)$ in the case $J(z)=e^{-|z|}$, $\beta=-1$, $c=1$, for the values
    $k=0.15$, $k=0.1$, $k=0.05$, $k=0.01$, and, in dashed line, $k=0$.}
    \label{sk0}
\end{figure}

We now present the results of some numerical computations of the
functions $g_k(\omega)$ for small values of $k$. In these we
encounter phenomenon not predicted by theorem \ref{smallkt}
(though of course it does not contradict it). We fix
$J(x)=e^{-|z|}$, $\beta=-1$ and $c=1$. In figure  \ref{sk0} we
plot the numerically-computed functions $g_k(\omega)$ for several
small values of $k$, as well as the function $g_0(\omega)$. We see
that as $k\rightarrow 0$. $g_k(\omega)$ converges to $g_0(\omega)$
uniformly on compact subsets of $(0,\infty)$. However we now see
that for sufficiently small $k$ the function $g_k$ has not one but
{\it{two}} minima. One of these minima approaches the minimum of
$g_0(\omega)$ as $k\rightarrow 0$, while the other one approaches
$0$. The observed shape of $g_k(\omega)$ shows that for some
values of $k$ and $g$ there exist {\it{four}} periodic waves of
wave-number $k$! It would be interesting to gain a better
understanding of the phenomena just described, both by means of
more systematic numerical investigations and if possible also by
analytical means.

\section{The question of stability}
\label{stab}

A crucial set of questions which are not addressed in our
investigations, and remain open for now, is the relevance of the
rotating waves on a ring, and of the periodic travelling waves on
a line, in terms of the full dynamical problem given by
(\ref{de}), (\ref{syn}). The general question, as with all
dynamical systems, is how arbitrary initial conditions develop in
the long-time limit. More particularly, we would like to know
whether the rotating waves on a ring and periodic travelling waves
on a line describe this asymptotic behavior, at least in some
cases.

A first step would be to determine the stability of the waves
whose existence was proved here as solutions of the full dynamical
problem. Let us note that the case of rotating wave on a ring and
of a periodic travelling wave on a line should be distinguished
here: if we can show that a rotating wave on a ring is
asymptotically stable, then it does {\it{not}} imply that the
corresponding periodic travelling wave on a line is stable - but
rather only that it is stable to perturbations which have the same
wave-number. Indeed, since in the case of a line, since we have a
continuum of possible periodic travelling waves for different
wave-numbers $k$, the best that we can expect is some kind of
`neutral stability' of the periodic travelling waves. If indeed
there is convergence (in some sense) to a periodic travelling
waves from general initial condition, the interesting question
arises as to how the wave number of limiting wave is selected.

Based on analytical numerical results on other models
\cite{bres,oe,osan,osan1}, we may conjecture that, at least under
some natural assumptions on the synaptic coupling kernels $K(x)$
(in the case of a ring), $J(x)$ (in the case of a line), the fast
wave is stable and the slow one is unstable. At our current state
of knowledge analytical results may be hard to obtain (but see
\cite{rubin} for some analytical progress on related stability
questions), so at least a systematic numerical investigation of
the full dynamics of (\ref{de}), (\ref{syn}) would be of much
interest.

\end{document}